\documentclass[twocolumn,showpacs,preprintnumbers,superscriptaddress,amsmath,amssymb,aps,pre]{revtex4-1}
\usepackage{epsfig}
\usepackage{graphicx}
\usepackage{dcolumn}
\usepackage{bm}
\usepackage{times}
\usepackage{xcolor}
\usepackage{booktabs}
\usepackage{subfigure}

\begin{document}

\title{Phase diagrams of interacting spreading dynamics in complex networks}

\author{Liming Pan}
\affiliation{School of Computer Science and Technology, Nanjing Normal University, Nanjing, Jiangsu, 210023, China}
\affiliation{Complex Lab, School of Computer Science and Engineering, University of Electronic Science and Technology of China, Chengdu, 611731, China}

\author{Dan Yang}
\affiliation{Complex Lab, School of Computer Science and Engineering, University of Electronic Science and Technology of China, Chengdu, 611731, China}

\author{Wei Wang}\email{wwzqbx@hotmail.com}
\affiliation{Cybersecurity Research Institute, Sichuan University, Chengdu 610065, China}
\affiliation{Complex Lab, School of Computer Science and Engineering, University of Electronic Science and Technology of China, Chengdu, 611731, China}

\author{Shimin Cai}
\affiliation{Complex Lab, School of Computer Science and Engineering, University of Electronic Science and Technology of China, Chengdu, 611731, China}
\affiliation{Institute of Fundamental and Frontier Sciences, University of Electronic Science and Technology of China, Chengdu 610054, China}
\affiliation{Big Data Research Center, University of Electronic Science and Technology of China, Chengdu 610054, China}

\author{Tao Zhou}
\affiliation{Complex Lab, School of Computer Science and Engineering, University of Electronic Science and Technology of China, Chengdu, 611731, China}
\affiliation{Institute of Fundamental and Frontier Sciences, University of Electronic Science and Technology of China, Chengdu 610054, China}
\affiliation{Big Data Research Center, University of Electronic Science and Technology of China, Chengdu 610054, China}

\author{Ying-Cheng Lai}
\affiliation{School of Electrical, Computer and Energy Engineering, Arizona State University, Tempe, AZ 85287, USA}

\date{\today}

\begin{abstract}
Epidemic spreading processes in the real world can interact with each other in a cooperative, competitive, or asymmetric way, requiring a description based on coevolution dynamics. Rich phenomena such as discontinuous outbreak transitions and hystereses can arise, but a full picture of these behaviors in the parameter space is lacking. We develop a theory for interacting spreading dynamics on complex networks through spectral dimension reduction. In particular, we derive from the microscopic quenched mean-field equations a two-dimensional system in terms of the macroscopic variables, which enables a full phase diagram to be determined analytically. The diagram predicts critical phenomena that were known previously but only numerically, such as the interplay between discontinuous transition and hysteresis as well as the emergence and role of tricritical points.
\end{abstract}

\maketitle

\section{Introduction} \label{sec:intro}

Spreading dynamics of diseases, behaviors and information in nature and human
society are rarely independent processes but interact with each other in a
complex way. Weakened immunity to other viruses due to HIV
infection~\cite{ferguson2003ecological,abu2006dual} and suppression of
spreading due to disease-related information exchange on the social
media~\cite{funk2009spread} are known examples. To better understand,
predict, and control spreading on networks, coevolution of epidemics must
be taken into account. In network science, there has been continuous
interest in developing interacting epidemic models~\cite{gog2002dynamics,
ferguson2003ecological,abu2006dual,EK:2006,berkman2014social,sanz2014dynamics,
pastor2015epidemic,de2018fundamentals,wang2019coevolution,SGMG:2019,
eames2006coexistence,danziger2019dynamic,soriano2019markovian},
which can generate surprising behaviors that cannot be predicted by any
single-virus epidemic model. For example, spreading of one epidemic can
facilitate that of another, leading to a first-order or explosive transition in
the outbreak with significant real-world implications~\cite{cai2015avalanche}.
Many factors can affect the critical behaviors of interacting spreading
dynamics, such as self-evolution of each epidemic~\cite{funk2009spread,
noh2005asymmetrically}, interaction between two
epidemics~\cite{granell2013dynamical,wang2014asymmetrically}, and
network structure~\cite{hebert2015complex,chen2019persistent}.

By now, spreading dynamics of a single epidemic on complex networks have been
well studied~\cite{castellano2009statistical,pastor2015epidemic,
dorogovtsev2008critical,kiss2017mathematics}. For interacting spreading
dynamics, the special case of well-mixed populations has been
treated~\cite{abu2008interactions,zarei2019exact}. A study based on the
quenched mean field for two competing pathogens~\cite{PBRF:2012} showed
that, when simultaneous infection by the two pathogens is not possible
(full mutual exclusion), the phase diagram is independent of the spectral
radius of the network. There were also theories based on
percolation~\cite{newman2005threshold}, annealed mean
field~\cite{chen2013outbreaks} and pair approximations~\cite{hebert2015complex}
to study the effect of network structure on interacting spreading, leading to 
a qualitative understanding of critical phenomena. There are difficulties with 
these theories. For example, the annealed mean-field theory takes into account 
only the nodal degrees and is not applicable to quenched networks (especially
networks with a high clustering coefficient and modularity).
For such networks, quenched mean-field theories~\cite{sahneh2014competitive}
such as those based on Markov chains~\cite{granell2014competing} and the
N-intertwined method~\cite{van2011n} is needed. A deficiency of such a
theory is that it uses a large number of nonlinear differential equations,
with two difficulties: (a) high computational overload for large networks
and (b) lack of any analytic insights. Such a theory, due to its heavy
reliance on numerics, can lead to inconsistent or even contradicting
predictions~\cite{ADS:2009,CDGM:2010}. To our knowledge, a general analytic
theory capable of providing a more complete understanding of interacting
spreading dynamics is lacking.

In this paper, we develop an analytic theory for interacting
spreading dynamics on complex networks through the approach of dimension
reduction for complex networks~\cite{gao2016universal,laurence2019spectral,
jiang2018predicting}. From the quenched mean-field equations, we derive a
two-dimensional (2D) system that is capable of analytically yielding the
full Phase diagrams underlying interacting spreading dynamics on any complex
network, from which the conditions for various phase transitions can be
derived. 
The analytic model predicts critical phenomena that were previously known 
numerically, such as the interplay between discontinuous outbreak transitions 
and hystereses as well as the emergence of tricritical points, providing 
a solid theoretical foundation for understanding interacting spreading 
dynamics and articulating optimal control strategies.

\section{Model, method of spectral dimension reduction, and reduced model} 
\label{sec:model}

\subsection{Model} \label{subsec:model}

We consider the susceptible-infected-susceptible (SIS) model of interacting
spreading dynamics on complex networks. In the classic SIS model, a single
epidemic spreads in the network and a node can be either in the susceptible or
in the infected state. Susceptible nodes are infected by their infected
neighbors at rate $\lambda$ and infected nodes recover at rate $\gamma$. For
interacting SIS dynamics, two epidemics, say $1$ and $2$, spread simultaneously
and interact with each other. Each node infected by $a\in \{1,2\}$ transmits
the infection to neighbors that are susceptible for both epidemics with
probability $\lambda_a$. If a neighbor is susceptible for $a$ but infected
by the other epidemic, the infection will be transmitted with rate
$\lambda^{\dagger}_a$. All the nodes infected by $a$ recover to being
susceptible with rate $\gamma_a$. Without loss of generality, we set
$\gamma_a=1$ for both $a\in\{1,2\}$. In general, the nature of the interacting
SIS dynamics depends on the interplay between the rates $\lambda_a$ and
$\lambda_a^{\dagger}$. In particular, for $\lambda_a^{\dagger}>\lambda_a$,
the two epidemics tend to facilitate each other, leading
to cooperative SIS dynamics, whereas if $\lambda_a^{\dagger}<\lambda_a$,
infection with one epidemic will suppress infection with
the other, giving rise to competitive SIS dynamics. For
$\lambda_a^{\dagger}>\lambda_a$ but $\lambda_b^{\dagger}<\lambda_b$ for
$a,b\in\{1,2\}$ with $\ b\neq a$, the interactions are asymmetric.

\subsection{Spectral dimension reduction} \label{sec:SDR}

The interacting SIS model represents a paradigm to study rich dynamical
behaviors such as first-order outbreak transitions and
hystereses~\cite{chen2017phase}. The foundation of our study of this model
is the quenched mean-field theory (QMF)~\cite{wang2003epidemic}. 
While in QMF the dynamical correlations among the neighbors are assumed to be
negligible, the theory has been demonstrated to generate reliable prediction
of the phase transitions~\cite{boguna2013nature}. Since our goal is to 
analytically map out the complete phase diagram, using the QMF suffices.
Let $p_{a,i}$ be the probability that node 
$i\in \{1,\cdots,N\}$ is infected
by $a\in \{1,2\}$ at time $t$. In the first-order mean-field
analysis~\cite{sahneh2014competitive}, the evolution of $p_{a,i}$ on a network
with adjacency matrix $G$ is governed by
\begin{equation} \label{eq:QMF}
\begin{split}
\frac{d p_{a,i}}{dt}=&-p_{a,i}+\lambda_a^{\dagger} \left(1-p_{a,i}\right)p_{b,i}\sum_j G_{ij}p_{a,j}\\
&+\lambda_a\left(1-p_{a,i}\right)\left(1-p_{b,i}\right)\sum_j G_{ij}p_{a,j}
\end{split}
\end{equation}
for $a\in\{1,2\}$ and $i\in\{1,\cdots,N\}$. The first term on the right side
of Eqs.~\eqref{eq:QMF} is the rate of recovery from epidemic $a$ for node $i$,
while the second (third) term corresponds to the rate of infection for epidemic
$a$ with (without) $i$ already infected by $b\neq a$. For a network of size
$N$, the number of equations in \eqref{eq:QMF} is $2N$. To derive an analytic
model, we exploit the technique of spectral dimension reduction
(SDR)~\cite{laurence2019spectral} to arrive at an equivalent description
of the original system in terms of two macroscopic observables - one for each
epidemic. In particular, let $\alpha$ be a vector with nonnegative entries
and normalized as $\sum_i \alpha_i=1$. The entries of $\alpha$ represent
the nodal weights. We define linear observables as $\psi_a=\alpha^{T} p_{a}$
for $a\in \{1,2\}$. Since the entries of $\alpha$ are summed to unity,
$\psi_a$ is a weighted average. The evolution of $\psi_a$ is determined by
the equation
\begin{equation} \label{eq:reducedQMF0}
\frac{d \psi_a}{dt}=\sum_{i=1}^{N}\alpha_i\frac{d p_{a,i}}{dt}.
\end{equation}
Applying the SDR method, we have that the right-hand side of
Eq.~\eqref{eq:reducedQMF0} can be written in terms of the macroscopic observables $\psi_a$ only.

\subsection{Reduced model} \label{sec:reduced_model}

We apply the SDR method to Eq.~\eqref{eq:reducedQMF0}. Microscopic variables 
$p_{a,i}$ fluctuate about the macroscopic observables $\psi_a$, which can
be decomposed as
\begin{equation} \label{eq:pDecomposition}
\begin{split}
&p_{a,i}=\rho_a \psi_a +\delta p_{a,i},\\
&p_{b,i}=\mu_a \psi_b +\delta p_{b,i},\\
&p_{a,j}=\nu_a \psi_a +\delta p_{a,j},
\end{split}
\end{equation}
where $\rho_a$, $\mu_a$, and $\nu_a$ are parameters to be determined, and
$\delta p_{a,i}$, $\delta p_{b,i}$, and $\delta p_{a,j}$ are correction terms.
Substituting Eqs.~\eqref{eq:QMF} and Eqs.~\eqref{eq:pDecomposition} into
Eqs.~\eqref{eq:reducedQMF0} gives
\begin{equation} \label{eq:mfObservable}
\begin{split}
\frac{d \psi_a}{dt}=&-\psi_a+\lambda^{\dagger}_a  \hat{\alpha} \mu_a\nu_a \left(1-\rho_a\psi_a\right)\psi_b \psi_a\\
&+\lambda_a\hat{\alpha} \nu_a \left(1-\rho_a\psi_a\right)(1-\mu_a\psi_b) \psi_a+R_a,
\end{split}
\end{equation}
where $\hat{\alpha}=\sum_{i,j}\alpha_i G_{ij}$ and $R_a$ is the remainder term
that can be decomposed as
\begin{equation}
R_a=R_{a,1}+R_{a,2}+R_{a,3},
\end{equation}
with $R_{a,1}$, $R_{a,2}$ and $R_{a,3}$ containing the first-, second- and
third-order terms in the corrections $\{\delta p_{a,i}\}$, respectively.
Let $K$ be the diagonal matrix with $K_{ii}$ being the degree of node $i$, the first-order correction $R_{a,1}$ is given by
\begin{equation} \label{eq:R1}
\begin{split}
&R_{a,1}=\left[\left(\lambda_a-\lambda_a^{\dagger}\right)\mu_a \nu_a \psi_b \psi_a-\lambda_a\nu_a\psi_a\right] \alpha^T K \delta p_a\\
&+\left(\lambda_a^{\dagger}-\lambda_a\right)\nu_a\left(1-\rho_a\psi_a\right)\psi_a\alpha^T K \delta p_b\\
& +\left(1-\rho_a\psi_a\right)(\lambda_a^{\dagger}\mu_a\psi_b+\lambda_a-\lambda_a\mu_a\psi_b)\alpha^T G \delta p_a,
\end{split}
\end{equation}
where $\delta p_a$ is a vector with $\delta p_{a,i}$ in the $i$th entry and
$\delta p_b$ is defined analogously. The second-order remainder term is
\begin{equation}
\begin{split}
&R_{a,2}=\left(\lambda_a^{\dagger}-\lambda_a\right)\left(1-\rho_a\psi_a\right)\sum_{i=1}^N \sum_{j=1}^N G_{ij} \alpha_i  \delta p_{b,i}\delta p_{a,j}\\
&+\left[\left(\lambda_a-\lambda_a^{\dagger}\right)\mu_a\psi_b-\lambda_a\right]\sum_{i=1}^N \sum_{j=1}^N G_{ij} \alpha_i \delta p_{a,i}\delta p_{a,j}\\
&+\left(\lambda_a-\lambda_a^{\dagger}\right)\nu_a\psi_a\sum_{i=1}^N \sum_{j=1}^N G_{ij} \alpha_i \delta p_{a,i}\delta p_{b,i}
\end{split}
\end{equation}
and the third-order remainder term is
\begin{equation}
R_{a,3}=\left(\lambda_a-\lambda_a^{\dagger}\right)\sum_{i=1}^N \sum_{j=1}^N G_{ij} \alpha_i \delta p_{a,i}\delta p_{b,i}\delta p_{a,j}.
\end{equation}
From Eq.~\eqref{eq:R1}, the dominant remainder term $R_{a,1}$ vanishes if
the following equations hold
\begin{equation} \label{eq:matrixEq}
\begin{split}
&\alpha^T K p_{a} = \hat{\alpha}\rho_a \psi_a,\\
&\alpha^T K p_b = \hat{\alpha} \mu_a \psi_b,\\
&\alpha^T G p_a =\hat{\alpha} \nu_a \psi_a,
\end{split}
\end{equation}
where $p_a$ is a vector with $p_{a,i}$ in the $i$th entry and $p_b$ is defined similarly. 

In general, the equations cannot be satisfied simultaneously. An application 
of the SDR method in Ref.~\cite{laurence2019spectral} advocates choosing 
$\alpha$ as the eigenvector associated with the leading eigenvalue $\omega$ 
of $G$. For connected undirected networks, the eigenvector associated with 
the leading eigenvalue $\omega$ of $G$ has positive entries. The third equation
in Eqs.~\eqref{eq:matrixEq} implies $\omega \psi_a=\hat{\alpha} \nu_a\psi_a$
and, hence, $\omega=\nu_a \hat{\alpha}$. Using the definition
$\hat{\alpha}=\mathbf{1}^T G \alpha = \omega$, we have $\nu_a=1$. The
remaining two parameters, $\rho_a$ and $\mu_b$, are chosen such that the
first two equations in Eqs.~\eqref{eq:matrixEq} are satisfied. The
quantities $\rho_a$ and $\mu_b$ can be chosen by minimizing the following
squared vector norm
\begin{equation}\nonumber
\begin{split}
\rho_a^*=\mu_a^*=\underset{x}{\mathrm{argmin}} \| K\alpha-x\hat{\alpha}\alpha \|_2^2,
\end{split}
\end{equation}
which yields
\begin{equation}\label{eq:muFormula}
\mu\mathop{:}=\frac{1}{\omega} \frac{\alpha^T K\alpha}{\alpha^T \alpha}=\rho_a^*=\mu_a^*.
\end{equation}
A justification of the parameter choices was given in
Ref.~\cite{laurence2019spectral}. With the parameters chosen, $R_{a,1}$
can be made as small as possible and can be neglected, so can the higher
order terms $R_{a,2}$ and $R_{a,3}$. Substituting the values of $\rho_a$,
$\mu_a$ and $\nu_a$ into Eqs.~\eqref{eq:mfObservable}, we get
\begin{equation} \label{eq:reducedQMF}
\begin{split}
\frac{d \psi_a}{dt}= &-\psi_a+\lambda^{\dagger}_a  \omega \mu\left(1-\mu \psi_a\right)\psi_b \psi_a\\
&+\lambda_a\omega \left(1-\mu \psi_a\right)(1-\mu\psi_b) \psi_a+R_a.
\end{split}
\end{equation}
The first term on the right side of Eqs.~\eqref{eq:reducedQMF} accounts for 
the rate of recovery and the second (third) term represents the rate of 
infection for epidemic $a$ with (without) being infected by $b\neq a$. The 
quantity $R_a$ in Eqs.~\eqref{eq:reducedQMF} characterizes the fluctuations 
of the microscopic observables $p_{a,i}$ about the macroscopic observables 
$\psi_a$, which is small in comparison to other terms on the right side of
Eqs.~\eqref{eq:reducedQMF} due to $\alpha$'s being the leading eigenvector. 
Since, for finding the phase diagram, it is necessary to analyze the 
mean-field equations that depend on the macroscopic observables $\psi_a$ only, 
it is justified to drop $R_a$ from the analysis. As we will verify numerically,
this approximation will not affect the accuracy of the phase diagram as the 
resulting errors near the phase boundaries are quite insignificant.

In Eqs.~\eqref{eq:QMF}, the order parameters are $\langle p_{a,i}\rangle$
for $a\in\{1,2\}$, where $\langle \cdot \rangle$ is the unweighted average
over the nodes. In Eqs.~\eqref{eq:reducedQMF}, the order parameters can be
chosen to be $\psi_a=\alpha^T p_{a}$, a weighted average over the nodes.
Since $\alpha$ is the eigenvector associated with the leading eigenvalue
of $G$, its entries are strictly positive. As a result, $\psi_a=0$
($\psi_a>0$) implies $\langle p_{a,i}\rangle=0$ ($\langle p_{a,i}\rangle>0$).
When crossing a phase boundary, at least one of $\psi_a$
for $a\in\{1,2\}$ becomes either zero or nonzero, guaranteeing that the
corresponding $\langle p_{a,i}\rangle$ becomes either zero or nonzero,
respectively. We have that $\langle p_{a,i}\rangle$ and $\psi_a$ give the
same phase diagram, which can be obtained analytically through the 2D
mean-field system.

\section{Main result: phase diagram of reduced system} \label{sec:PD}

The reduced mean-field equations are amenable to analytic treatment. As the 
derivations are lengthy, we provide a brief sketch of the results from 
analyzing the reduced system.

The analyses of the 2D mean-field system are performed in the following steps. 
First, for each point in the parameter space
$(\lambda_1$, $\lambda_1^{\dagger},\lambda_2$, $\lambda_2^{\dagger})$,
we determine the equilibrium points of Eqs.~\eqref{eq:reducedQMF} 
(Sec.~\ref{subsec:EPs}) and their stability (Sec.~\ref{subsec:SA}). The
the equilibrium points have to further satisfy the probability constraint
$1\leq \psi_a\leq 1$ to be physical meaningful. A detailed analysis of the
stability and probability constraints of the equilibrium points leads to the
following functions of $\lambda_a$ and $\lambda_a^{\dagger}$:
\begin{equation}
\begin{split}
s_{a,0}=&\lambda_b+\lambda^{\dagger}_a-\lambda_a-\omega\lambda_b\lambda^{\dagger}_a,\\
s_{a,1}=&\lambda^{\dagger}_b-\lambda_b-\lambda^{\dagger}_a+\lambda_a  +2\omega\lambda_b\lambda^{\dagger}_a-\omega \lambda^{\dagger}_a\lambda^{\dagger}_b,\\
s_{a,2}=&\lambda^{\dagger}_b-\lambda_b,\ s_{a,3}=\lambda_a-\omega^{-1},\\
s_{\Delta}=&(\lambda^{\dagger}_1-\lambda_1 +\lambda^{\dagger}_2-\lambda_2- \omega\lambda^{\dagger}_1\lambda^{\dagger}_2)^2,\\
&-4(\lambda^{\dagger}_1-\lambda_1)(\lambda^{\dagger}_2-\lambda_2).
\end{split}
\end{equation}
for $a,b \in \{1,2\}$ and $a\neq b$. Whether an equilibrium point is physical 
or stable is determined by the signs of these functions. 

Calculating the equilibrium points (Sec.~\ref{subsec:EPs}) and analyzing 
their stability (Sec.~\ref{subsec:SA}) enable us to obtain the full phase 
diagram of the reduced system and the equations for the phase boundaries 
(Sec.~\ref{subsec:phaseDiagram}). Based on the numbers of stable and unstable 
equilibrium points as well as the relationships among them, we can divide the 
parameter space into distinct regions, where a boundary crossing between two 
neighboring regions gives rise to a phase transition. A region either can have 
a unique stable equilibrium point or can have two stable equilibrium points 
with one unstable point in between, where crossing the latter will result in a 
hysteresis. We discuss the types of phase transitions crossing the various 
boundaries and study the interplay between the transitions and the phenomenon
of hysteresis (Sec.~\ref{subsec:TPT}). Finally, we derive the conditions under 
which a hysteresis can arise  (Sec.~\ref{subsec:hysteresis}). 

The analytical results of the full phase diagram are summarized and discussed 
in Sec.~\ref{subsec:SoP}. Readers who are not interested in the technical 
details of analyzing the 2D mean-field system can skip Secs.~\ref{subsec:EPs} 
to \ref{subsec:hysteresis} and check Sec.~\ref{subsec:SoP} for the results.
For convenience, for the rest of the paper, we use the convention that, if
variables indexed by $a,b\in\{1,2\}$ (e.g., $\psi_a$ and  $\psi_b$) appear
together in an equation or an inequality, the assumption is $a\neq b$.

\subsection{Equilibrium points of the reduced system} \label{subsec:EPs}

The equilibrium points are obtained by setting the right side of
Eqs.~\eqref{eq:reducedQMF} to zero:
\begin{equation} \label{eq:equilibrium}
\begin{split}
&-\psi_a+\lambda^{\dagger}_a  \omega \mu \left(1-\mu\psi_a\right)\psi_b \psi_a\\
&+\lambda_a\omega \left(1-\mu\psi_a\right)(1-\mu\psi_b) \psi_a =0
\end{split}
\end{equation}
for $a,b\in\{1,2\}$ and $a\neq b$. Further, the physical solutions have to
satisfy the probability constraints $0\leq\psi_a\leq 1$. Because of the
appearance of terms such as $\left(1-\mu \psi_a\right)$ in
Eqs.~\eqref{eq:reducedQMF}, it is necessary to impose the physical
condition $\left(1-\mu \psi_a\right)\leq 1$. It can be proved that $\mu$ 
given by Eq.~\eqref{eq:muFormula} satisfies $\mu\geq 1$ (see Appendix A for 
a proof), and it can also be verified that any point with $\psi_1=\mu^{-1}$ 
or $\psi_2=\mu^{-1}$ cannot be an equilibrium point. These, together the 
probability constraints, imply that all the equilibrium points must satisfy 
the inequality $0\leq\psi_a< \mu^{-1}$ for $a\in \{1,2\}$.

We are now in a position to discuss the types of equilibrium points of the
reduced mean-field equations.

\noindent (\romannumeral1) \textit{Epidemic free.} The trivial solution
$\left(\psi_1,\psi_2\right)=(0,0)$ is always an equilibrium point.

\noindent (\romannumeral2) \textit{Partial infection of epidemic $1$}.
For $\psi_1 \neq 0$ and $\psi_2=0$, Eqs.~\eqref{eq:equilibrium} become
\begin{equation}
-1+\lambda_1\omega \left(1-\mu\psi_1\right)=0,
\end{equation}
which gives
\begin{equation}
\psi_1=\frac{\lambda_1 \omega-1}{\mu\lambda_1 \omega}.
\end{equation}
The solution further has to satisfy the probability constraint
$0<\psi_1\leq \mu^{-1}$. The first inequality $\psi_1>0$ is satisfied when 
$\lambda_1\omega>1$, while the second inequality $\psi_1\leq \mu^{-1}$ always 
holds.

\noindent (\romannumeral3) \textit{Partial infection of epidemic $2$}.
Similar to case (\romannumeral2), we have the equilibrium point:
\begin{equation}
\left(\psi_1,\psi_2\right)=\left(0,\frac{\lambda_2 \omega-1}{\mu\lambda_2 \omega}\right).
\end{equation}

\noindent (\romannumeral4) \textit{Coexistence}.
If $\psi_a \neq 0$ for both $a\in \{1,2\}$, Eqs.~\eqref{eq:equilibrium} become
\begin{subequations} \label{eq:equilibrium2}
\begin{align}
\lambda^{\dagger}_1  \omega \mu \left(1-\mu \psi_1\right)\psi_2+\lambda_1\omega \left(1-\mu \psi_1\right)(1-\mu \psi_2) =1,\\
\lambda^{\dagger}_2  \omega \mu \left(1-\mu \psi_2\right)\psi_1+\lambda_2\omega \left(1-\mu \psi_2\right)(1-\mu \psi_1) =1.
\end{align}
\end{subequations}
Rearranging the second equation, we get
\begin{equation}
\psi_2=\mu^{-1}-\frac{1}{\left(\lambda^{\dagger}_2-\lambda_2\right)  \mu^2 \omega \psi_1 + \lambda_2\mu\omega}.
\end{equation}
Substituting this relation into Eq.~(\ref{eq:equilibrium2}a), we get an
equation that depends on $\psi_1$ only. Similarly we can obtain the equation
that determines $\psi_2$. The two equations for $\psi_1$ and $\psi_2$ have
the following symmetric form:
\begin{equation} \label{eq:psiEquation}
g_{a,2} \psi_a^2+g_{a,1}\psi_a+g_{a,0}=0
\end{equation}
for $a\in\{1,2\}$, where
\begin{equation} \label{eq:gDef}
\begin{split}
g_{a,2}&=\omega^2 \mu^3\lambda^{\dagger}_a\left(\lambda^{\dagger}_b-\lambda_b\right),\\
g_{a,1}&=\omega \mu^2\left(\lambda^{\dagger}_b-\lambda_b-\lambda^{\dagger}_a+\lambda_a  +2\omega\lambda_b\lambda^{\dagger}_a-\omega \lambda^{\dagger}_a\lambda^{\dagger}_b\right),\\
g_{a,0}&=  \omega \mu \left(\lambda_b+\lambda^{\dagger}_a-\lambda_a-\omega\lambda_b\lambda^{\dagger}_a\right)
\end{split}
\end{equation}
for $b\in \{1,2\}$ and $b\neq a$.

If $\lambda_a^{\dagger}\neq \lambda_a$ holds for $a\in \{1,2\}$,
$g_{a,2}\neq 0$ and Eqs.~\eqref{eq:psiEquation} will be quadratic, leading
to two solutions
\begin{equation} \label{eq:coinfectionSolution}
\psi_a^{\pm}=\frac{-g_{a,1}\pm \sqrt{g_{a,1}^2-4g_{a,2}g_{a,0}}}{2g_{a,2}}.
\end{equation}
The solutions for $a\in\{1,2\}$ are paired as
\begin{equation}\label{eq:psiPair}
\left(\psi_1,\psi_2\right)=\left(\psi_1^+,\psi_2^+\right),\ \left(\psi_1,\psi_2\right)=\left(\psi_1^-,\psi_2^-\right).
\end{equation}

If $\lambda_a^{\dagger}=\lambda_a$ for one or both $a\in \{1,2\}$, we have
\begin{equation}
\psi_a=-\frac{g_{a,0}}{g_{a,1}}.
\end{equation}

To discuss the probability constraints of the equilibrium points, we consider
the following cases.

\noindent (\romannumeral4 .1) If $g_{a,2}=0$ for one of $a\in\{1,2\}$,
i.e., $\lambda_b^{\dagger}=\lambda_b$, then there is a unique solution
given by
\begin{equation}
\psi_a=\mu^{-1}-\frac{\lambda_b}{\mu\left(\lambda_a-\lambda_a^{\dagger}+\omega\lambda_b\lambda_a^{\dagger}\right)},\ \psi_b=\mu^{-1}-\frac{1}{\mu\omega\lambda_b}.
\end{equation}
The probability constraints imply
\begin{equation}
\lambda_a-\lambda_b-\lambda_a^{\dagger}+\omega\lambda_b\lambda_a^{\dagger}>0,\ \omega\lambda_b>1.
\end{equation}
Further, if we have $\lambda_a^{\dagger}=\lambda_a$, the two epidemics will
become independent of each other with the solution
\begin{equation}
\psi_a=\mu^{-1}-\frac{1}{\omega\mu\lambda_a},\ \psi_b=\mu^{-1}-\frac{1}{\omega\mu\lambda_b}.
\end{equation}

\noindent (\romannumeral4 .2) Suppose $\lambda_a^{\dagger}>\lambda_a$ for
both $a\in \{1,2\}$. In this case there are two solutions, as shown in
Eqs.~\eqref{eq:psiPair}.

Consider the solution
\begin{displaymath}
\left(\psi_1,\psi_2\right)=\left(\psi_1^+,\psi_2^+\right).
\end{displaymath}	
Firstly, it is necessary to have $g_{a,1}^2-4g_{a,2}g_{a,0}\geq 0$ for
$a\in \{1,2\}$ to make the solutions real. Because of the condition
$g_{a,2}>0$, the probability constraints $0<\psi_a<\mu^{-1}$ imply
\begin{equation} \label{eq:pbConstraint1}
g_{a,1}<\sqrt{g_{a,1}^2-4g_{a,2}g_{a,0}}< 2\mu^{-1} g_{a,2}+g_{a,1}
\end{equation}
for $a\in\{1,2\}$. The second inequality can be written as
\begin{equation}
\mu^{-2}g_{a,2}+\mu^{-1}g_{a,1}+g_{a,0}> 0.
\end{equation}
Substituting these into Eqs.~\eqref{eq:gDef}, we have
\begin{equation}
g_{a,0}+\mu^{-1}g_{a,1}+\mu^{-2}g_{a,2}=\mu \omega\lambda_b^{\dagger}>0,
\end{equation}
indicating that the second inequality always holds.

It remains to consider the first inequality in Eqs.~\eqref{eq:pbConstraint1}.
Suppose $g_{a,0}<0$, then both the first and the inequality
$g_{a,1}^2-4g_{a,2}g_{a,0}\geq 0$ hold. Otherwise, suppose
$g_{a,0}>0$, it is necessary to have $g_{a,1}<0$ and
$g_{a,1}^2-4g_{a,2}g_{a,0}\geq 0$.

Combining the discussions above, we have that $\left(\psi_1^+,\psi_2^+\right)$ 
is a physical solution either for $g_{a,0}<0$ or for $g_{a,0}>0$, $g_{a,1}<0$, 
$g_{a,1}^2-4g_{a,2}g_{a,0}\geq 0$ for both $a\in\{1,2\}$.

We now consider the solution
$\left(\psi_1,\psi_2\right)=\left(\psi_1^-,\psi_2^-\right)$. In order for the
solution to be meaningful, it has to be guaranteed that
$g_{a,1}^2-4g_{a,2}g_{a,0}\geq 0$ for $a\in \{1,2\}$. The probability
constraints give
\begin{equation}
g_{a,1}<-\sqrt{g_{a,1}^2-4g_{a,2}g_{a,0}}< 2\mu^{-1} g_{a,2}+g_{a,1}.
\end{equation}
The first inequality implies $g_{a,1}<0$ and $g_{a,0}>0$. Consider the
second inequality. If $2\mu^{-1} g_{a,2}+g_{a,1}> 0$, then the second
inequality will be satisfied. Else if $2\mu^{-1} g_{a,2}+g_{a,1}\leq 0$,
the second inequality can be  written as
\begin{equation}
\mu^{-2}g_{a,2}+\mu^{-1}g_{a,1}+g_{a,0}\leq 0.
\end{equation}
Substituting these into Eqs.~\eqref{eq:gDef} we have
\begin{equation}
g_{a,0}+\mu^{-1}g_{a,1}+\mu^{-2}g_{a,2}=\mu \omega\lambda_b^{\dagger}>0,
\end{equation}
which leads to a contradiction. It is thus necessary to have
$2\mu^{-1} g_{a,2}+g_{a,1}\geq 0$ for $a\in \{1,2\}$. In fact, the
inequalities $g_{a,1}<0$ and $g_{a,0}>0$ are sufficient to guarantee
the condition $2\mu^{-1} g_{a,2}+g_{a,1}\geq 0$. For $g_{1,1}+g_{2,1}<0$,
we have
\begin{equation}
\lambda^{\dagger}_1+\lambda^{\dagger}_2-\omega \lambda^{\dagger}_1\lambda^{\dagger}_2<0.
\end{equation}
We then have
\begin{equation}
\begin{split}
2\mu^{-1} g_{1,2}+g_{1,1}=&\omega \mu^2\left(\lambda^{\dagger}_2-\lambda_2-\lambda^{\dagger}_1+\lambda_1+\omega \lambda^{\dagger}_1\lambda^{\dagger}_2\right)\\
>&\omega \mu^2\left(2\lambda^{\dagger}_2-\lambda_2+\lambda_1\right)>0.
\end{split}
\end{equation}
Similarly, we obtain $2\mu^{-1} g_{2,2}+g_{2,1}>0$.

Combining the conditions discussed above, we have that the solution
$\left(\psi_1,\psi_2\right)=\left(\psi_1^-,\psi_2^-\right)$ is physical
for $g_{a,2}>0$, $g_{a,1}<0$, $g_{a,0}>0$ and $g_{a,1}^2-4g_{a,2}g_{a,0}>0$
for $a\in\{1,2\}$. Comparing with the conditions for
$\left(\psi_1,\psi_2\right)=\left(\psi_1^+,\psi_2^+\right)$, we see that,
for $g_{a,0}<0$, only one physical solution is possible.

\noindent (\romannumeral4 .3)
Suppose $\lambda_a^{\dagger}<\lambda_a$ for $a\in \{1,2\}$ and
$\lambda_b^{\dagger}>\lambda_b$ for $b\in \{1,2\}$ and $b\neq a$. We have
$g_{a,2}>0$ and $g_{b,2}<0$. Consider the solution
$\left(\psi_1,\psi_2\right)=\left(\psi_1^+,\psi_2^+\right)$. The probability
constraints imply
\begin{subequations} \label{eq:localLabel1}
\begin{align}
&g_{a,1}<\sqrt{g_{a,1}^2-4g_{a,2}g_{a,0}}< 2\mu^{-1} g_{a,2}+g_{a,1},\\
&g_{b,1}>\sqrt{g_{b,1}^2-4g_{b,2}g_{b,0}}> 2\mu^{-1} g_{b,2}+g_{b,1}.
\end{align}
\end{subequations}
From Eq.~(\ref{eq:localLabel1}a) we must have either $g_{a,0}<0$ or
$g_{a,0}>0$, $g_{a,1}<0$, $g_{a,1}^2-4g_{a,2}g_{a,0}>0$. The first inequality
in Eq.~(\ref{eq:localLabel1}b) implies $g_{b,1}>0$ and $g_{b,0}<0$. Now
consider the second inequality in Eq.~(\ref{eq:localLabel1}b). For
$2\mu^{-1} g_{b,2}+g_{b,1}\leq 0$, the second inequality in
Eq.~(\ref{eq:localLabel1}b) holds. Otherwise if $2\mu^{-1} g_{b,2}+g_{b,1}>0$,
the second inequality implies
\begin{equation}
c_{b,0}+\mu^{-1}c_{b,1}+\mu^{-2}c_{b,2}>0
\end{equation}
which always holds since the left side of the above inequality equals
$\mu \omega\lambda_a^{\dagger}$.

Recall that, from Eq.~(\ref{eq:localLabel1}a), we can have either $g_{a,0}<0$
or $g_{a,0}>0$, $g_{a,1}<0$, $g_{a,1}^2-4g_{a,2}g_{a,0}>0$. We can show that
the latter case contradicts with the conditions $g_{b,1}>0$ and $g_{b,0}<0$.
In particular, from
\begin{equation}
g_{b,0}=\omega\mu\left(\lambda_a\left(1-\omega\lambda_b^{\dagger}\right)+\lambda_b^{\dagger}-\lambda_b\right)<0,
\end{equation}
we have $1-\omega\lambda_b^{\dagger}<0$ and similarly
\begin{equation}
g_{a,0}=\omega\mu\left(\lambda_b\left(1-\omega\lambda_a^{\dagger}\right)+\lambda_a^{\dagger}-\lambda_a\right)>0,
\end{equation}
implying $1-\omega\lambda_a^{\dagger}>0$. Since
\begin{equation}
\begin{split}
g_{a,1}&=\omega \mu^2\left(\lambda^{\dagger}_b-\lambda_b-\lambda^{\dagger}_a+\lambda_a  +2\omega\lambda_b\lambda^{\dagger}_a-\omega \lambda^{\dagger}_a\lambda^{\dagger}_b\right)\\
&=\omega \mu^2\left(\left(\lambda_b^{\dagger}-\lambda_b\right)\left(1-\omega\lambda_a^{\dagger}\right)+\lambda_a-\lambda_a^{\dagger}\left(1-\omega\lambda_b\right)\right),
\end{split}
\end{equation}
then $g_{a,1}<0$ implies
\begin{equation}
\lambda_a-\lambda_a^{\dagger}\left(1-\omega\lambda_b\right)<0.
\end{equation}
As a result, we have
$\left(1-\omega\lambda_b\right)>\lambda_a/ \lambda_a^{\dagger}>1$
and $\omega\lambda_b<0$, leading to a contradiction.

Summarizing the above discussions about the equilibrium points, we have
that $\left(\psi_1,\psi_2\right)=\left(\psi_1^+,\psi_2^+\right)$ is physical
for $g_{a,0}<0$, $g_{b,0}<0$ and $g_{b,1}>0$. Note that $g_{b,1}>0$ is
implied by the other two. Since
\begin{equation}
g_{b,1}+\mu g_{b,0}=\omega \mu^2\left(\lambda_a^{\dagger}+\omega\lambda_b^{\dagger}\left(\lambda_a-\lambda_a^{\dagger}\right)\right)>0,
\end{equation}
we have that $g_{b,1}>0$ always holds given $g_{b,0}<0$. Together, it is
sufficient to have $g_{a,0}<0$ and $g_{b,0}<0$.

We consider the solution
$\left(\psi_1,\psi_2\right)=\left(\psi_1^-,\psi_2^-\right)$. The probability
constraints imply
\begin{subequations} \label{eq:localLabel2}
\begin{align}
&g_{a,1}<-\sqrt{g_{a,1}^2-4g_{a,2}g_{a,0}}< 2\mu^{-1} g_{a,2}+g_{a,1},\\
&g_{b,1}>-\sqrt{g_{b,1}^2-4g_{b,2}g_{b,0}}> 2\mu^{-1} g_{b,2}+g_{b,1}.
\end{align}
\end{subequations}
From Eq.~(\ref{eq:localLabel2}b) we have $2\mu^{-1} g_{b,2}+g_{b,1}<0$, giving
\begin{equation}
g_{b,0}+\mu^{-1}g_{b,1}+\mu^{-2}g_{b,2}< 0.
\end{equation}
which cannot hold since its left side equals
$\mu \omega\lambda_a^{\dagger}$. Thus, in this region, no physical solution
of $\left(\psi_1,\psi_2\right)=\left(\psi_1^-,\psi_2^-\right)$ exists.

\noindent (\romannumeral4 .4)
Suppose $\lambda_a^{\dagger}<\lambda_a$ for $a\in \{1,2\}$, then
$c_{a,2}<0$. For the solution
$\left(\psi_1,\psi_2\right)=\left(\psi_1^+,\psi_2^+\right)$, we must have
$g_{a,1}>0$ and $g_{a,0}<0$ for $a\in \{1,2\}$. Similar to the discussions
in the case (\romannumeral4 .3), we have that the sufficient condition for
an equilibrium point is $g_{a,0}<0$ for $a\in \{1,2\}$. The solution
$\left(\psi_1,\psi_2\right)=\left(\psi_1^-,\psi_2^-\right)$ is nonphysical
- see the discussion in (\romannumeral4 .3).

\subsection{Stability analysis} \label{subsec:SA}

The starting point to study the stability of the equilibrium points is the
Jacobian matrix $J$ of the 2D mean-field system, whose entries are
\begin{equation} \label{eq:Jacobian}
\begin{split}
J_{11}=&-1+\lambda^{\dagger}_1  \omega \mu \left(1-2x\psi_1\right)\psi_2\\
&+\lambda_1  \omega \left(1-2\mu\psi_1\right)(1-\mu\psi_2), \\
J_{12}=&\omega \mu\left(\lambda^{\dagger}_1-\lambda_1\right) \left(1-\mu\psi_1\right)\psi_1,\\
J_{21}=&\omega \mu\left(\lambda^{\dagger}_2-\lambda_2\right) \left(1-\mu\psi_2\right)\psi_2,\\
J_{22}=&-1+\lambda^{\dagger}_2  \omega \mu \left(1-2\mu\psi_2\right)\psi_1\\
&+\lambda_2  \omega \left(1-2\mu\psi_2\right)(1-\mu\psi_1).
\end{split}
\end{equation}
We analyze the stability of the different classes of equilibrium points
as discussed in Sec.~\ref{subsec:EPs}.

\noindent (\romannumeral1) \textit{Epidemic free}. For
$\left(\psi_1,\psi_2\right)=(0,0)$, the Jacobian matrix is
\begin{equation}
J=\left(
{\begin{array}{cc}
-1+\lambda_1 \omega & 0 \\
0 & -1+\lambda_2 \omega
\end{array}}
\right),
\end{equation}
The equilibrium point is stable for $\lambda_a<\omega^{-1}$ for $a\in\{1,2\}$.

\noindent (\romannumeral2) \textit{Partial infection of epidemic $1$}. In
this case, we have
\begin{equation}
\left(\psi_1,\psi_2\right)=\left(\frac{\lambda_1 \omega-1}{\mu\lambda_1 \omega},0\right)
\end{equation}
and $J_{21}=0$, so the Jacobian is upper triangular, whose eigenvalues are
simply the diagonal entries:
\begin{equation}
\begin{split}
J_{1,1}&=1-\lambda_1 \omega\\
J_{2,2}&=-1+\frac{\lambda_2}{\lambda_1}+\lambda_2^{\dagger}\frac{\lambda_1 \omega-1}{\lambda_1 }
\end{split}
\end{equation}
The equilibrium point is stable for $J_{1,1}<0$ and $J_{2,2}<0$, i.e.,
\begin{equation}
\begin{split}
&\lambda_1>\omega^{-1},\\
&\lambda_2-\lambda_2^{\dagger}-\lambda_1+\omega\lambda_1\lambda_2^{\dagger}<0.
\end{split}
\end{equation}

\noindent (\romannumeral3) \textit{Partial infection of epidemic $2$}.
For this type of equilibrium point, we have
\begin{equation}
\left(\psi_1,\psi_2\right)=\left(0,\frac{\lambda_2 \omega-1}{\mu\lambda_2 \omega}\right).
\end{equation}
It is stable under the following conditions:
\begin{equation}
\begin{split}
&\lambda_2>\omega^{-1},\\
&\lambda_1-\lambda_1^{\dagger}-\lambda_2+\omega\lambda_2\lambda_1^{\dagger}<0.
\end{split}
\end{equation}

\noindent (\romannumeral4) \textit{Coexistence}.
Suppose we have $\psi_a \neq 0$ for $a\in \{1,2\}$. Substituting
Eqs.~\eqref{eq:equilibrium2} into Eqs.~\eqref{eq:Jacobian}, the Jacobian matrix has entries
\begin{equation}
\begin{split}
&J_{11}=\mu\psi_1/(\mu\psi_1-1),\\
&J_{12}=\omega \mu \left(\lambda^{\dagger}_1-\lambda_1\right)  \left(1-\mu\psi_1\right)\psi_1,\\
&J_{21}=\omega \mu\left(\lambda^{\dagger}_2-\lambda_2\right)  \left(1-\mu\psi_2\right)\psi_2,\\
&J_{22}=\mu\psi_2/(\mu\psi_2-1).
\end{split}
\end{equation}
A necessary and sufficient condition for a two-dimensional matrix to have
two negative eigenvalues is to have a negative trace ($\mathrm{tr} (J)<0$)
but a positive determinant ($\mathrm{det} (J)>0$). Since $\mu \psi_a-1<0$,
the negativity of the trace always holds. The stability of a equilibrium
point in this class is fully determined by the determinant. It is stable when
$\mathrm{det} (J)>0$ and unstable when $\mathrm{det} (J)>0$. The stable
condition from the determinant is
\begin{equation}
\begin{split}
\mathrm{det} (J)=&\frac{\mu^2\psi_a \psi_b}{(1-\mu\psi_a)(1-\mu\psi_b)}+\\
&- \omega^2 \mu^2\left(\lambda^{\dagger}_a-\lambda_a\right)\left(\lambda^{\dagger}_b-\lambda_b\right) \\
&\times \left(1-\mu\psi_a\right)\left(1-\mu\psi_b\right)\psi_a \psi_b >0.
\end{split}
\end{equation}
Let $z=\left(1-\mu\psi_a\right)\left(1-\mu\psi_b\right)$, the inequality
can be written as
\begin{equation} \label{eq:detIneq}
\frac{1}{z}> \omega^2 \left(\lambda^{\dagger}_a-\lambda_a\right)\left(\lambda^{\dagger}_b-\lambda_b\right)z.
\end{equation}
Equations~\eqref{eq:equilibrium2} can be rearranged as
\begin{subequations}
\begin{align}
&\lambda^{\dagger}_1  \omega\left(1-\mu \psi_1\right)=1+\left(\lambda_1^{\dagger}-\lambda_1\right)\omega z,\\
&\lambda^{\dagger}_2  \omega\left(1-\mu \psi_2\right)=1+\left(\lambda_2^{\dagger}-\lambda_2\right)\omega z.
\end{align}
\end{subequations}
Multiplying the above two equations, we get
\begin{equation} \label{eq:pairEquation}
d_2 z^{-2}+d_1z^{-1}+d_0=0,
\end{equation}
where
\begin{equation}
\begin{split}
&d_2=1,\\
&d_1=\omega\left(\lambda^{\dagger}_1-\lambda_1 +\lambda^{\dagger}_2-\lambda_2-  \omega\lambda^{\dagger}_1\lambda^{\dagger}_2\right),\\
&d_0=\omega^2\left(\lambda^{\dagger}_1-\lambda_1 \right)\left(\lambda^{\dagger}_2-\lambda_2 \right).
\end{split}
\end{equation}
Multiplying both sides of Eq.~\eqref{eq:pairEquation} by $z$ and substituting
the result into Eq.~\eqref{eq:detIneq}, we obtain
\begin{equation} \label{eq:detIneq2}
\frac{1}{z}>-\frac{d_1}{2}.
\end{equation}
That is, an equilibrium point is stable if and only if Eq.~\eqref{eq:detIneq2}
holds and is unstable otherwise. It remains to find the solutions of
Eq.~\eqref{eq:pairEquation} to verify whether Eq.\eqref{eq:detIneq2} is
satisfied.

If the condition $\lambda_b^{\dagger}=\lambda_b$ holds for one or both
values of $b\in \{1,2\}$, then $d_0=0$. In this case, we have $z^{-1}=-d_1/d_2$
and Eq.~\eqref{eq:detIneq} implies the solution is stable for $d_1<0$.

For $\lambda_a^{\dagger}\neq \lambda_b$ for any $a\in \{1,2\}$, from
Eq.~\eqref{eq:pairEquation}, we see that $1/z$ has two solutions
\begin{equation}\label{eq:zSolution}
\left(\frac{1}{z}\right)^{\pm}=\frac{-d_1\pm\sqrt{d_1^2-4d_2d_0}}{2d_2}.
\end{equation}
Since we have a pair of solutions for $\left(\psi_1,\psi_2\right)$ as in
Eq.~\eqref{eq:psiPair}, the following hold:
\begin{equation}
\begin{split}
&\left(\frac{1}{z}\right)^{+}=\frac{1}{\left(1-\mu\psi_a^+\right)\left(1-\mu\psi_b^+\right)},\\
& \left(\frac{1}{z}\right)^{-}=\frac{1}{\left(1-\mu\psi_a^-\right)\left(1-\mu\psi_b^-\right)}.
\end{split}
\end{equation}
Substituting Eq.~\eqref{eq:zSolution} into Eq.\eqref{eq:detIneq2}, we have
\begin{equation}
\pm\sqrt{d_1^2-4d_2d_0}\geq 0.
\end{equation}
We see that, given $d_1^2-4d_2d_0>0$, the solution
$\left(\psi_a^+,\psi_b^+\right)$ is always stable, while
$\left(\psi_a^-,\psi_b^-\right)$ is always unstable. It remains to check
the validity of the inequality $d_1^2-4d_2d_0>0$. After some algebra, we have
\begin{equation}
\begin{split}
&d_1^2-4d_2d_0=g_{1,1}^2-4g_{1,2}g_{1,0}=g_{2,1}^2-4g_{2,2}g_{2,0}\\
=&\omega\mu^2\left(\lambda^{\dagger}_1-\lambda_1 +\lambda^{\dagger}_2-\lambda_2- \omega\lambda^{\dagger}_1\lambda^{\dagger}_2\right)^2+\\
&-4\omega\mu^2\left(\lambda^{\dagger}_1-\lambda_1 \right)\left(\lambda^{\dagger}_2-\lambda_2 \right).
\end{split}
\end{equation}
Thus the inequalities $d_1^2-4d_2d_0>0$ and $g_{a,1}^2-4g_{a,2}g_{a,0}>0$
are equivalent to each other for $a\in\{1,2\}$.

\subsection{Phase diagrams} \label{subsec:phaseDiagram}

With full knowledge about the equilibrium points and their stability, we
can obtain the phase diagram of the reduced mean-field system. Define the
following set of functions
\begin{equation} \label{eq:sDef}
\begin{split}
s_{a,0}= &\lambda_b+\lambda^{\dagger}_a-\lambda_a-\omega\lambda_b\lambda^{\dagger}_a,\\
s_{a,1}=&\lambda^{\dagger}_b-\lambda_b-\lambda^{\dagger}_a+\lambda_a  +2\omega\lambda_b\lambda^{\dagger}_a-\omega \lambda^{\dagger}_a\lambda^{\dagger}_b,\\
s_{a,2}=&\lambda^{\dagger}_b-\lambda_b,\ 
s_{a,3}=\lambda_a-\omega^{-1},\\
s_{\Delta}=&\left(\lambda^{\dagger}_1-\lambda_1 +\lambda^{\dagger}_2-\lambda_2- \omega\lambda^{\dagger}_1\lambda^{\dagger}_2\right)^2\\
&-4\left(\lambda^{\dagger}_1-\lambda_1 \right)\left(\lambda^{\dagger}_2-\lambda_2 \right).
\end{split}
\end{equation}
for $a,b \in \{1,2\}$ and $a\neq b$. The distinct phase regions can be
defined via various inequalities among these functions.

\noindent (\romannumeral1) \textit{Epidemic free}. The solution
$\left(\psi_1,\psi_2\right)=(0,0)$ is stable for $s_{a,3}<0$ for both
$a\in\{1,2\}$.

\noindent (\romannumeral2) \textit{Partial infection of epidemic $1$}.
The phase has a stable equilibrium point
\begin{equation}
\left(\psi_1,\psi_2\right)=\left(\frac{\lambda_1 \omega-1}{\mu\lambda_1 \omega},0\right).
\end{equation}
Combining the probability constraints and the stability analysis, we obtain
the phase region as
\begin{equation}
s_{2,0}>0,\ s_{1,3}>0.
\end{equation}

\noindent (\romannumeral3) \textit{Partial infection of epidemic $2$}.
The phase is characterized by
\begin{equation}
\left(\psi_1,\psi_2\right)=\left(0,\frac{\lambda_2 \omega-1}{\mu\lambda_2 \omega}\right).
\end{equation}
The phase region is given by
\begin{equation}
s_{1,0}>0,\ s_{2,3}>0.
\end{equation}

\noindent (\romannumeral4) \textit{Coexistence}.
In this region, there is an equilibrium point with both $\psi_1$ and $\psi_2$
nonzero, corresponding to the case of double epidemic outbreaks. For
cooperative coevolution, i.e., $\lambda_a^{\dagger}>\lambda_a$ for
$a\in\{1,2\}$, a point in the parameter space belongs to this phase if
\begin{equation}
s_{a,0}>0,\ s_{a,1}<0,\ s_{a,2}>0, \ s_{\Delta}>0,
\end{equation}
or
\begin{equation}
s_{a,0}<0, \ s_{a,2}>0
\end{equation}
for both $a\in\{1,2\}$. When coevolution is not cooperative, the coexistence
region is given by
\begin{equation}
s_{a,0}<0
\end{equation}
for both $a\in\{1,2\}$. We have verified that the case of
$\lambda_a=\lambda_a^{\dagger}$ for one or both $a\in\{1,2\}$ is well
covered by this inequality.

\noindent (\romannumeral1 $\ \cap$ \romannumeral4).
\textit{Hysteresis region 1}. A hysteresis region appears when there are
two stable equilibrium points and one unstable equilibrium point in between.
The stability analysis indicates that the solution
$\left(\psi_1^{+},\psi_2^{+}\right)$ is always stable while
$\left(\psi_1^{-},\psi_2^{-}\right)$ is unstable. In addition to these two
equilibrium points, a third stable solution is necessary for a hysteresis
to arise. This is only possible when region (\romannumeral4) overlaps with
regions (\romannumeral1), (\romannumeral2) and (\romannumeral3). Checking
the equilibrium points and their stability, we find that a hysteresis region
exists only when the inequality $\lambda_a^{\dagger}>\lambda_a$ holds for
$a\in\{1,2\}$. The region where (\romannumeral1) and (\romannumeral4)
overlap is
\begin{equation}
s_{a,0}>0,\ s_{a,1}<0,\ s_{a,2}>0,\ s_{a,3}<0,\ s_{\Delta}>0,
\end{equation}
where the first inequality $s_{a,0}>0$ can in fact be implied by the
other inequalities. Since $g_{1,1}+g_{2,1}<0$, we have
\begin{equation}
\lambda^{\dagger}_1+\lambda^{\dagger}_2-\omega \lambda^{\dagger}_1\lambda^{\dagger}_2<0,
\end{equation}
which further implies $\omega\lambda^{\dagger}_1>1$ and
$\omega\lambda^{\dagger}_2>1$. Since $s_{a,3}<0$, we have
\begin{equation}
s_{a,0}=\lambda_b\left(1-\omega\lambda^{\dagger}_a\right)+\lambda^{\dagger}_a-\lambda_a>\omega^{-1}-\lambda_a>0.
\end{equation}
Altogether, the region is given by
\begin{equation}
\ s_{a,1}<0,\ s_{a,2}>0,\ s_{a,3}<0,\ s_{\Delta}>0
\end{equation}
for $a\in\{1,2\}$.

\noindent (\romannumeral2 $\ \cap$ \romannumeral4).
\textit{Hysteresis region 2}. This region is where (\romannumeral2) and
(\romannumeral4) overlap, which is bounded by the inequalities
\begin{equation}
s_{a,0}>0,\ s_{a,1}<0,\ s_{a,2}>0,\ s_{1,3}>0,\ s_{2,3}<0,\ s_{\Delta}>0
\end{equation}
for $a\in\{1,2\}$.

\noindent (\romannumeral3$\ \cap$ \romannumeral4).
\textit{Hysteresis region 3}. Similarly, the region where (\romannumeral3)
and (\romannumeral4) overlap is bounded by
\begin{equation}
s_{a,0}>0,\ s_{a,1}<0,\ s_{a,2}>0,\ s_{1,3}<0,\ s_{2,3}>0,\ s_{\Delta}>0
\end{equation}
for $a\in\{1,2\}$.

\subsection{Types of phase transition} \label{subsec:TPT}

A phase transition occurs when a point in the parameter space crosses a
boundary between two neighboring phase regions. Depending on different
combinations of phase-region pairs, the resulting phase transitions can be
characteristically distinct. To be concrete, we focus on the phase transitions
in the $\lambda_1$-$\lambda_2$ plane with fixed values of $\lambda_1^{\dagger}$
and $\lambda_2^{\dagger}$. Both continuous and discontinuous phase transitions
can arise, as we will show below.

\noindent (\romannumeral1) $\rightleftharpoons$ (\romannumeral2):
We have that the equations $s_{1,0}=0$ and $s_{2,0}=0$ intersect at the point
$\left(\lambda_1,\lambda_2\right)=\left(\omega^{-1},\omega^{-1}\right)$,
so the two phases are separated by the line $s_{1,3}=0$ in the
$\lambda_1$-$\lambda_2$ plane. When approaching the line $s_{1,3}=0$ from
phase (\romannumeral2), the equilibrium point
\begin{equation}
\left(\psi_1,\psi_1\right)=\left(\frac{\lambda_1 \omega-1}{\mu\lambda_1 \omega},0\right)
\end{equation}
approaches $\left(\psi_1,\psi_1\right)=(0,0)$. As a result, a continuous
phase transition arises.

\noindent (\romannumeral1) $\rightleftharpoons$ (\romannumeral3): Similar
to the preceding case, the phase transition is continuous.

\noindent (\romannumeral2) $\rightleftharpoons$ (\romannumeral4)$\setminus$(\romannumeral2 $\ \cap$ \romannumeral4):
The two phases are separated by the line $s_{2,0}=0$. When the stable
equilibrium point $\left(\psi_1^+,\psi_2^+\right)$ in
Eqs.~\eqref{eq:coinfectionSolution} approaches the line, for $s_{2,1}>0$
we have
\begin{equation}
\left(\psi_1^+,\psi_2^+\right)\rightarrow \left(\frac{\lambda_1 \omega-1}{\mu\lambda_1 \omega},0\right),
\end{equation}
generating a continuous phase transition. Otherwise ($s_{2,1}<0$), we have
\begin{equation}\label{eq:locaLabel4}
\left(\psi_1^+,\psi_2^+\right)\rightarrow \left(-\frac{g_{1,2}+g_{2,1}}{g_{1,2}},-\frac{g_{2,1}}{g_{2,2}}\right),
\end{equation}
so the phase transition is discontinuous. It remains to discuss the sign of
$s_{2,1}$. Substituting $s_{2,0}=0$ into $s_{2,1}$, we get
\begin{equation}\label{eq:localLabel3}
s_{2,1}=\lambda_1^{\dagger}+\omega \lambda_1\lambda_2^{\dagger}-\omega \lambda_1^{\dagger}\lambda_2^{\dagger}.
\end{equation}
First consider the case where the coevolution dynamics are not cooperative,
i.e., $\lambda_a\geq \lambda_a^{\dagger}$ for at least one of $a\in\{1,2\}$.
In this case, the region (\romannumeral2 $\ \cap$ \romannumeral4) is empty.
Suppose $\lambda_1\geq \lambda_1^{\dagger}$, it can be immediately seen
that $s_{2,1}>0$. For $\lambda_2\geq \lambda_2^{\dagger}$, we have $s_{2,0}>0$,
implying $\omega_2\lambda_2^{\dagger}\leq 1$ and consequently $s_{2,1}>0$.

Now consider the case of cooperative coevolution dynamics, where a point in the
region (\romannumeral2 $\ \cap$ \romannumeral4) satisfies $s_{2,1}<0$.
Further, we can prove that, if a point is in the region
(\romannumeral4)$\setminus$(\romannumeral2 $\ \cap$ \romannumeral4), then
$s_{2,1}>0$. This is accomplished by showing that if a point has $s_{2,1}<0$
then it must be in the region (\romannumeral2 $\ \cap$ \romannumeral4).
Notice that the equations $s_{2,1}=0$, $s_{s_{2,0}}=0$ and $s_{\Delta}=0$
intersect at the point
\begin{equation} \label{eq:tricritical1}
\left(\lambda_1,\lambda_2\right)=\left(\lambda_1^{\dagger}-\frac{\lambda_1^{\dagger}}{\omega\lambda_2^{\dagger}},2\lambda_1^{\dagger}+\lambda_2^{\dagger}-\omega\lambda_1^{\dagger}\lambda_2^{\dagger}-\frac{\lambda_1^{\dagger}}{\omega\lambda_2^{\dagger}}\right)
\end{equation}
in the $\lambda_1$-$\lambda_2$ plane. Since $s_{2,1}$ is an increasing
function of $\lambda_1$ along $s_{2,0}=0$, as can be seen from
Eq.~\eqref{eq:localLabel3}, we have that, if a point in the line defined
by $s_{2,0}=0$ in the $\lambda_1$-$\lambda_2$ plane has
$\lambda_1<\lambda_1^{\dagger}-\lambda_1^{\dagger}/\omega\lambda_2^{\dagger}$,
it will satisfy $s_{2,1}<0$. Furthermore, since $\lambda_1>\omega^{-1}$, the
inequality
$\lambda_1<\lambda_1^{\dagger}-\lambda_1^{\dagger}/\omega\lambda_2^{\dagger}$ implies
\begin{equation}
\lambda_1^{\dagger}+\lambda_2^{\dagger}<\omega\lambda_1^{\dagger}\lambda_2^{\dagger}.
\end{equation}
Along the line $s_{2,0}=0$, $s_{1,0}$ can be written as
\begin{equation}
s_{1,0}=\left(\omega\lambda_1^{\dagger}\lambda_2^{\dagger}-\lambda_1^{\dagger}-\lambda_2^{\dagger}\right)\lambda_1+\lambda_1^{\dagger}+\lambda_2^{\dagger}-\omega\lambda_1^{\dagger}\lambda_2^{\dagger},
\end{equation}
which is an increasing function of $\lambda$. Since the curves $s_{1,0}=0$
and $s_{2,0}=0$ intersect at the point
$\left(\lambda_1,\lambda_2\right)=\left(\omega^{-1},\omega^{-1}\right)$,
we have $s_{1,0}=0$. We thus have $s_{1,0}>0$ for $\lambda_1>\omega^{-1}$.
Similarly, along the line $s_{2,0}=0$, we have
\begin{equation}
\begin{split}
s_{1,1}&=-\lambda_1^{\dagger}+\omega\lambda_1\lambda_2^{\dagger}+2\omega\lambda_2\lambda_1^{\dagger}-\omega\lambda_1^{\dagger}\lambda_2^{\dagger}\\
&<-2\lambda_1^{\dagger}+2\omega\lambda_2\lambda_1^{\dagger}<0.
\end{split}
\end{equation}
The first inequality is the result of
$\lambda_1<\lambda_1^{\dagger}-\lambda_1^{\dagger}/\omega\lambda_2^{\dagger}$ and the second inequality is due to the fact $\lambda_2<\omega^{-1}$ along
the line $s_{2,0}=0$ for $\lambda_1>\omega^{-1}$. Lastly, a point in
region (\romannumeral2) can always make $s_{\Delta}>0$ if it is sufficiently
close to the line $s_{2,0}=0$.

To summarize, if a point in region (\romannumeral2) has $s_{2,1}<0$ near
the phase boundary $s_{2,0}=0$, then all the conditions under which the point
is in region (\romannumeral2 $\ \cap$ \romannumeral4) hold. Thus, if the point
is in the region
(\romannumeral4)$\setminus$(\romannumeral2 $\ \cap$ \romannumeral4), we
have $s_{2,1}>0$, which makes the phase transition continuous.

\noindent (\romannumeral3) $\rightleftharpoons$ (\romannumeral4)$\setminus$(\romannumeral3 $\ \cap$ \romannumeral4).
Following a similar treatment to the preceding case, we have that the phase
transition is continuous.

\noindent (\romannumeral2 $\ \cap$ \romannumeral4)$\rightarrow$(\romannumeral4).
The two phase are separated by the line $s_{2,0}=0$. As discussed in the
case of the (\romannumeral2) $\rightleftharpoons$ (\romannumeral4)$\setminus$(\romannumeral2 $\ \cap$ \romannumeral4) transition, since $s_{2,1}<0$ holds
near the phase boundary, the behavior of the coexistence solution is
determined by Eq.~\eqref{eq:locaLabel4} when approaching the phase boundary,
resulting in a discontinuous phase transition.

With discussions similar to those in the
(\romannumeral2 $\ \cap$ \romannumeral4) $\rightarrow$ (\romannumeral4) case,
we find that all transitions as a result of entering or leaving the
hysteresis region are of the discontinuous type, due to the fact that, in
the hysteresis region, the inequality $s_{a,1}<0$ holds. The discontinuous
transitions include (\romannumeral2 $\ \cap$ \romannumeral4) $\rightarrow$ (\romannumeral4), (\romannumeral3 $\ \cap$ \romannumeral4) $\rightarrow$ (\romannumeral4), (\romannumeral1 $\ \cap$ \romannumeral4) $\rightarrow$ (\romannumeral4), (\romannumeral2 $\ \cap$ \romannumeral4) $\rightarrow$ (\romannumeral2)$\setminus$(\romannumeral2 $\ \cap$ \romannumeral4), (\romannumeral3 $\ \cap$ \romannumeral4) $\rightarrow$ (\romannumeral3)$\setminus$(\romannumeral3 $\ \cap$ \romannumeral4) and (\romannumeral1 $\ \cap$ \romannumeral4) $\rightarrow$ (\romannumeral1)$\setminus$(\romannumeral1 $\ \cap$ \romannumeral4).

The tricritical points that separate the continuous from the discontinuous
transition lie in the boundaries of the hysteresis region where $q_{a,1}=0$
holds for either $a\in\{1,2\}$. One such point is given by
Eq.~\eqref{eq:tricritical1}. The second tricritical point can be obtained
similarly as
\begin{equation} \label{eq:tricritical2}
\left(\lambda_1,\lambda_2\right)=\left(2\lambda_2^{\dagger}+\lambda_1^{\dagger}-\omega\lambda_1^{\dagger}\lambda_2^{\dagger}-\frac{\lambda_2^{\dagger}}{\omega\lambda_1^{\dagger}},\lambda_2^{\dagger}-\frac{\lambda_2^{\dagger}}{\omega\lambda_1^{\dagger}}\right).
\end{equation}

\subsection{Conditions on $\lambda_a^{\dagger}$ for hysteresis}
\label{subsec:hysteresis}

For fixed values of $\lambda_1^{\dagger}$ and $\lambda_2^{\dagger}$, a
hysteresis can arise in the $\lambda_1$-$\lambda_2$ plane. To determine these
values, we first note that a hysteresis is possible only when the coevolution
dynamics are cooperative, i.e., $\lambda_a^{\dagger}>\lambda_a$ for both
$a\in\{1,2\}$. A point in the hysteresis region must satisfy the inequality
$g_{1,1}+g_{2,1}<0$. Consequently, we have
\begin{equation} \label{eq:hysteresisCondL}
\lambda^{\dagger}_1+\lambda^{\dagger}_2-\omega \lambda^{\dagger}_1\lambda^{\dagger}_2<0,
\end{equation}
which provides a necessary condition for a hysteresis to arise. We can show
that this is also sufficient to guarantee the occurrence of a hysteresis.
In particular, suppose inequality Eq.~\eqref{eq:hysteresisCond} holds.
Since $s_{1,0}=0$ and $s_{2,0}=0$ intersect at
$\left(\lambda_1,\lambda_2\right)=\left(\omega^{-1},\omega^{-1}\right)$ in
the $\lambda_a$-$\lambda_b$ plane, there is a neighborhood near
$\left(\omega^{-1},\omega^{-1}\right)$ in which the inequalities
$s_{1,0}>0$ and $s_{2,0}>0$ hold. At the point
$(\lambda_a,\lambda_b)=\left(\omega^{-1},\omega^{-1}\right)$, we have
\begin{equation}
s_{1,1}=s_{2,1}=\lambda^{\dagger}_1+\lambda^{\dagger}_2-\omega \lambda^{\dagger}_1\lambda^{\dagger}_2<0.
\end{equation}
It remains to check whether the inequality $s_{\Delta}>0$ holds. Let
$\omega \lambda^{\dagger}_a\lambda^{\dagger}_b=\lambda^{\dagger}_a+\lambda^{\dagger}_b+\epsilon$,
where $\epsilon>0$ is a constant. Then at the point
$(\lambda_a,\lambda_b)=\left(\omega^{-1},\omega^{-1}\right)$, we have
\begin{equation}
s_{\Delta}=\left(2\omega^{-1}+\epsilon\right)^2-4\left(\epsilon\omega^{-1}+\omega^{-2}\right)=\epsilon>0.
\end{equation}
We thus have that a hysteresis region exists in the $\lambda_1$-$\lambda_2$
plane if and only if Eq.~\eqref{eq:hysteresisCondL} holds.

\subsection{Summary of the phase diagrams}\label{subsec:SoP}

Based on the results of the above analysis, we obtain the structure of the 
analytically predicted phase diagram, which can be described, as follows.

\paragraph*{(\romannumeral1) Epidemic free region.}
In this region, Eqs.~\eqref{eq:reducedQMF} has the equilibrium point
$\left(\psi_1,\psi_2\right)=(0,0)$, indicating extinction of both epidemics.
The solution is stable for $s_{a,3}<0$ ($a\in\{1,2\}$). The phase boundary
$\lambda_a=\omega^{-1}$ is also the outbreak threshold for the classic SIS
model with a single epidemic.

\paragraph*{(\romannumeral2) Partial infection of epidemic $1$.}
In this phase, 
an outbreak occurs for the epidemic $1$ but not for $2$.
The equilibrium point is given by
\begin{equation}
(\psi_1,\psi_2)=\left(\frac{\lambda_1 \omega-1}{\mu\lambda_1 \omega},0\right),
\end{equation}
which is stable for $s_{2,0}>0$ and $s_{1,3}>0$, where the latter
gives $\lambda_1>\omega^{-1}$, indicating that epidemic $1$ can have an
outbreak independently. Similarly, $s_{2,0}>0$ implies $\lambda_2<\lambda_1$
and
\begin{equation}
\lambda_2^{\dagger} < (\lambda_1-\lambda_2)/(\omega\lambda_1-1),
\end{equation}
stipulating that $\lambda_2^{\dagger}$ cannot be too large, such that the 
outbreak of epidemic $1$ result in an outbreak in epidemic $2$.

\paragraph*{(\romannumeral3) Partial infection of epidemic $2$.}
Analogous to {\em (\romannumeral2)}, this phase is defined by
\begin{equation}
\left(\psi_1,\psi_2\right)=\left(0,\frac{\lambda_2 \omega-1}{\mu\lambda_2 \omega}\right),
\end{equation}
which is stable for $s_{1,0}>0$ and $s_{2,3}>0$.

\paragraph*{(\romannumeral4) Coexistence.}
In this region the reduced system has a stable equilibrium point with both
$\psi_1$ and $\psi_2$ nonzero, leading to a simultaneous outbreak of two
epidemics. The stable equilibrium points are
\begin{equation} \label{eq:coinfectionSolutionL}
\left(\psi_1,\psi_2\right)=\left(\frac{-s_{1,1}\pm \sqrt{s_{\Delta}}}{2\mu \omega\lambda_1^{\dagger} s_{1,2}},\frac{-s_{2,1}\pm \sqrt{s_{\Delta}}}{2\mu \omega\lambda_2^{\dagger} s_{2,2}}\right).
\end{equation}
For cooperative coevolution, i.e., $\lambda_a^{\dagger}>\lambda_a$ for
$a\in\{1,2\}$, we have $s_{a,2}>0$. A point in the parameter space belongs
to this phase if it further satisfies either
\begin{equation}
s_{a,0}>0, \ s_{a,1}<0, \ s_{\Delta}>0,
\end{equation}
or
\begin{equation}
s_{a,0}<0
\end{equation}
for $a\in\{1,2\}$. The former case is where region {\em (\romannumeral4)}
overlaps with regions {\em (\romannumeral1)}, {\em (\romannumeral2)} and
{\em (\romannumeral3)}. As a result, hystereses can arise. For competitive 
or asymmetric coevolution, the coexistence region is given by $s_{a,0}<0$ 
for $a\in\{1,2\}$.

Now we discuss the conditions for observing the coexistence phase in more 
detail to get an intuitive picture. For the cooperative case, the boundaries 
of the coexistence region are relatively complex, and we consider the 
degenerate case of 
$\lambda_1=\lambda_2$ and $\lambda_1^{\dagger}=\lambda_2^{\dagger}$.
For $s_{a,0}<0$, we have $\omega \lambda_1=\omega \lambda_2>1$ so that
both epidemics are able to have an outbreak by themselves. On the contrary, for
$s_{a,0}>0$ so that neither epidemic can have an outbreak by itself, we have
$s_{a,1}<0$ and $\lambda_1^{\dagger}>2\lambda_1$. The condition $s_{\Delta}>0$
requires
\begin{equation}
(\omega\lambda_1^{\dagger})^2>4(\omega\lambda_1^{\dagger}-\omega\lambda_1),
\end{equation}
and $\lambda_1^{\dagger}>2\lambda_1$, leading to the necessary condition
$\omega\lambda_1^{\dagger}>2$. In this case, the interaction transmission
rate must at least double the threshold value of classic SIS outbreak
to have coexistence.

For the competitive case, since $s_{a,0}<0$, we have
\begin{equation}
s_{1,0}+s_{2,0}=\lambda_1^{\dagger}\left(1-\omega\lambda_2\right)+\lambda_2^{\dagger}\left(1-\omega\lambda_2\right)<0.
\end{equation}
The above inequality implies that at least one of $\lambda_a$, say $\lambda_1$, must satisfy $\omega \lambda_1>1$. Actually, we must also have $\omega \lambda_2>1$ to observe the coexistence phase. Suppose $\omega \lambda_2<1$, from $\omega s_{1,0}<0$, we have
\begin{equation}
\begin{split}
&\omega\lambda_1-\omega\lambda_2+\omega \lambda_2^{\dagger}\left(1-\omega\lambda_1\right)<0\\
\Rightarrow\ &\omega \lambda_2^{\dagger}>\frac{\omega\lambda_2-\omega\lambda_1}{1-\omega\lambda_1}>1>\omega \lambda_2,
\end{split}
\end{equation}
which contradicts with the assumption that the model is competitive. 
Consequently, to observe the coexistence of two epidemics for the 
competitive case, a necessary condition is that each of the two epidemics 
can have an outbreak by itself. In addition, $s_{a,0}<0$ implies 
\begin{equation}\label{eq:localLabel5}
\lambda_a^{\dagger}>\frac{\lambda_a-\lambda_b}{1-\omega \lambda_b}.
\end{equation}
That is to say, to have the coexistence phase, the suppression effect from 
the other epidemic cannot be too strong. For networks with a larger value of 
$\omega$, the conditions Eq.~\eqref{eq:localLabel5} and $\omega\lambda_a>1$ 
both can be readily satisfied. As a result, the coexistence phase is more 
likely to be observed in networks with a larger leading eigenvalue $\omega$.

It is also interesting to note that, when the two epidemics have fully mutual 
exclusion (i.e., $\lambda_a^{\dagger}=0$), the condition $s_{a,0}<0$ can never 
be satisfied simultaneously for both $a\in \{1,2\}$. In other words, the 
coexistence phase can never be observed with fully mutual exclusion, and this 
phenomenon agrees with the prediction in Ref.~\cite{prakash2012winner}.

\paragraph*{(\romannumeral1 $\ \cap$ \romannumeral4) Hysteresis region 1.}
A hysteresis arises when there are two stable equilibrium
points and one unstable equilibrium point in between, which occurs when
region {\em (\romannumeral4)} overlaps with regions {\em (\romannumeral1)},
{\em (\romannumeral2)} and {\em (\romannumeral3)}. Our analysis reveals
that a hysteresis region emerges only for cooperative coevolution.
The region where {\em (\romannumeral1)} and {\em (\romannumeral4)} overlap
is bounded by the inequalities
\begin{equation}
s_{a,1}<0,\ s_{a,2}>0,\ s_{a,3}<0,\ s_{\Delta}>0
\end{equation}
for $a\in\{1,2\}$.

\paragraph*{(\romannumeral2 $\ \cap$ \romannumeral4) Hysteresis region 2.}
This is where {\em (\romannumeral2)} and {\em (\romannumeral4)} overlap.
Besides the cooperative condition $s_{a,2}>0$, the region is nonempty if
the inequalities
\begin{equation}
s_{a,0}>0,\ s_{a,1}<0,\ s_{1,3}>0,\ s_{2,3}<0,\ s_{\Delta}>0
\end{equation}
hold for $a\in\{1,2\}$.

\paragraph*{(\romannumeral3$\ \cap$ \romannumeral4) Hysteresis region 3.}
Similarly, for cooperative coevolution with $s_{a,2}>0$, the region where
$(\romannumeral3)$ and $(\romannumeral4)$ overlap is bounded by
\begin{equation}
s_{a,0}>0,\ s_{a,1}<0,\ s_{1,3}<0,\ s_{2,3}>0,\ s_{\Delta}>0
\end{equation}
for $a\in\{1,2\}$.

The types of phase transitions that occur when crossing a phase boundary 
are determined by further checking if the stable solution varies continuously 
(see Sec.~\ref{subsec:TPT}). We find all possible phase transitions as a 
result of crossing a hysteresis region are discontinuous, whereas other 
transitions are continuous. 
A result revealed by our analysis of the phase diagrams is that the precursor 
of a discontinuous transition with an abrupt outbreak of at least one epidemic 
is a hysteresis.
Continuous and discontinuous phase transitions are separated by two
tricritical points in the $\lambda_1$-$\lambda_2$ plane:
\begin{equation} \label{eq:tricritical}
\begin{split}
&(\lambda_1,\lambda_2)=\left(\lambda_1^{\dagger}-\frac{\lambda_1^{\dagger}}{\omega\lambda_2^{\dagger}}, 2\lambda_1^{\dagger}+\lambda_2^{\dagger}-\omega\lambda_1^{\dagger}\lambda_2^{\dagger}-\frac{\lambda_1^{\dagger}}{\omega\lambda_2^{\dagger}}\right),\\
&(\lambda_1,\lambda_2)=\left(2\lambda_2^{\dagger}+\lambda_1^{\dagger}-\omega\lambda_1^{\dagger}\lambda_2^{\dagger} -\frac{\lambda_2^{\dagger}}{\omega\lambda_1^{\dagger}},\lambda_2^{\dagger}-\frac{\lambda_2^{\dagger}}{\omega\lambda_1^{\dagger}}\right).
\end{split}
\end{equation}
The phase diagram also makes it possible to obtain the conditions in the
interaction strengths $\lambda_1^{\dagger}$ and $\lambda_2^{\dagger}$ for
a hysteresis to occur. In Sec.~\ref{subsec:hysteresis}, we obtain the necessary
and sufficient condition
\begin{equation} \label{eq:hysteresisCond}
\lambda^{\dagger}_1+\lambda^{\dagger}_2-\omega \lambda^{\dagger}_1\lambda^{\dagger}_2<0,
\end{equation}
where there is a hysteresis region with $\lambda_1<\lambda_1^{\dagger}$
and $\lambda_2<\lambda_2^{\dagger}$.

\section{Numerical verification} \label{sec:numerics}

\begin{figure*} [ht!]
\centering
\includegraphics[width=\linewidth]{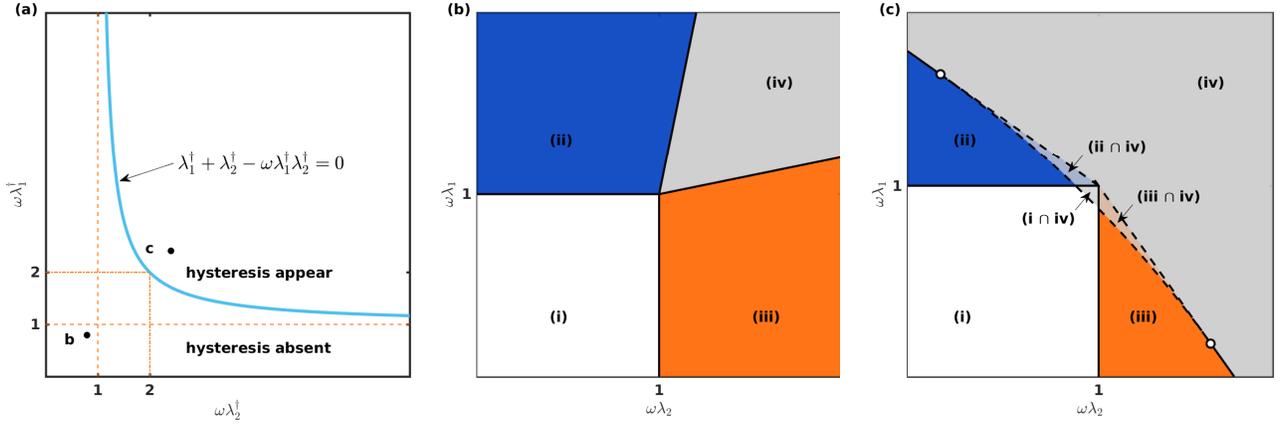}
\caption{ {\emph{ Phase diagrams of interacting SIS dynamics}}.
For Erd\H{o}s-R\'{e}nyi type of random graph of size $N=100$ and average
degree $\langle k \rangle=4$, (a) the condition for interactive transmission
rates $\lambda_1^{\dagger}$ and $\lambda_2^{\dagger}$ for hysteresis in the
$\lambda_1$-$\lambda_2$ plane. The red dashed and dashed-dotted lines
correspond, respectively, to $\lambda_a=\omega^{-1}$ and
$\lambda_a=2\omega^{-1}$ for $a\in\{1,2\}$. (b,c) Phase diagrams
with $\lambda_1^{\dagger}$ and $\lambda_2^{\dagger}$ corresponding to
points $b$ and $c$ in (a), respectively. The solid and dashed lines between
different phase regions indicate continuous and discontinuous
transitions, respectively. The two white circles in (c) are the tricritical
points separating discontinuous from continuous transitions.}
\label{fig:PD}
\end{figure*}

\begin{figure*} [ht!]
\centering
\subfigure{\includegraphics[width=0.49\linewidth]{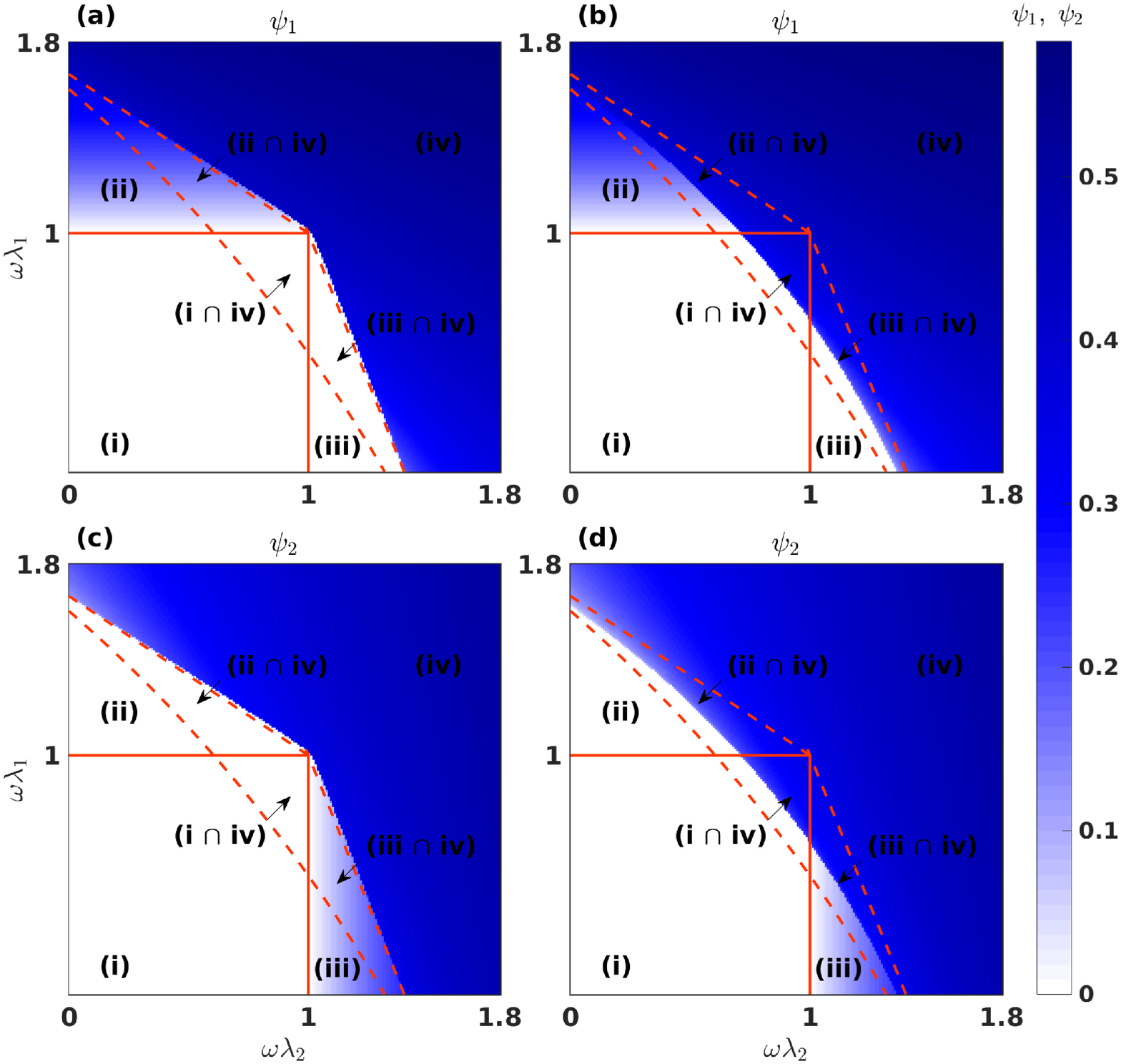}}
\subfigure{\includegraphics[width=0.49\linewidth]{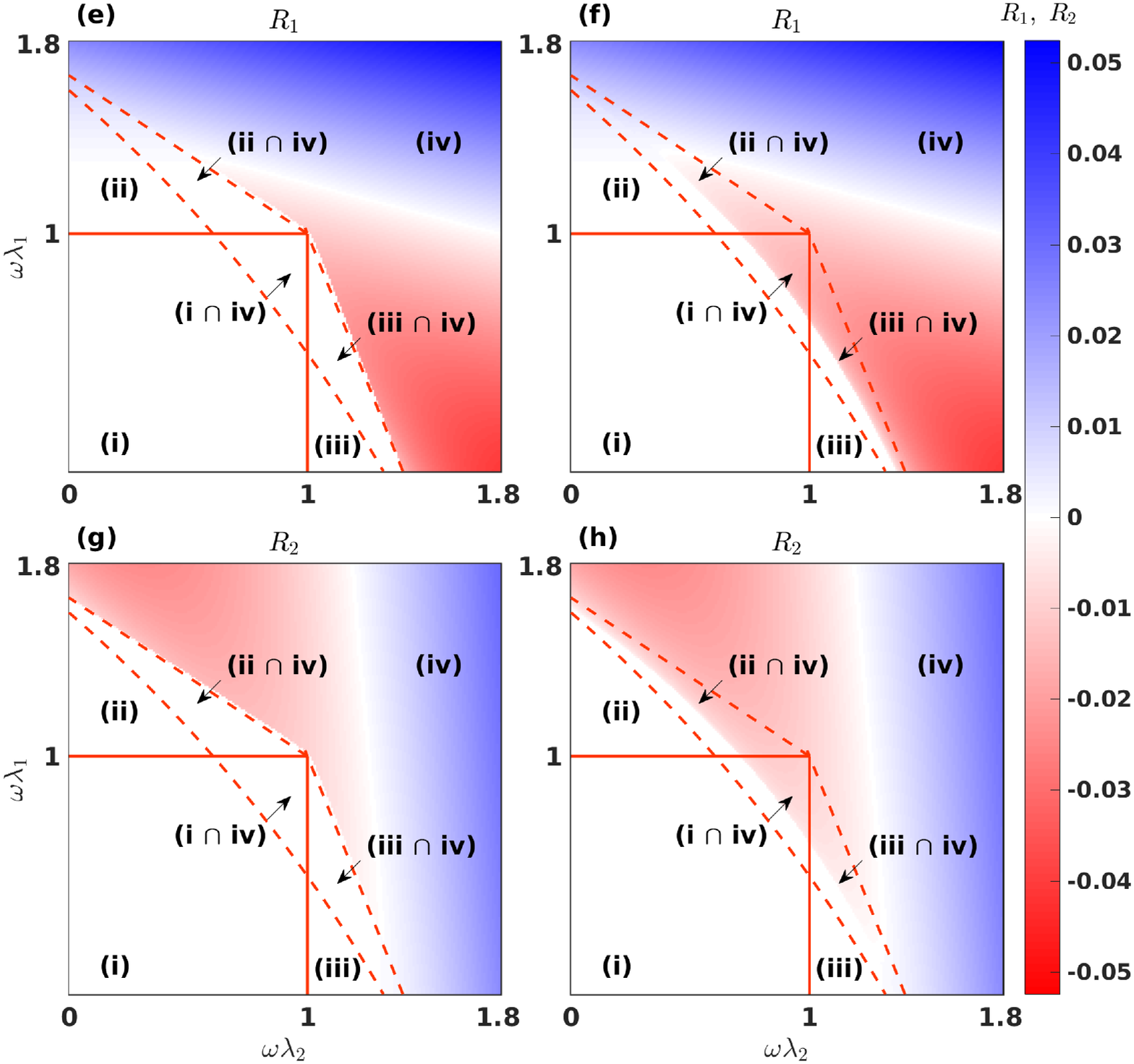}}
\caption{ \emph{Validation of analytic predictions}. Shown is comparison 
of the analytically predicted phase diagram with that obtained from the 
original mean field equations: equilibrium points for color-coded 
(a,b) $\psi_1$ and (c,d) $\psi_2$ values; color-coded values of the 
remainder terms $R_1$ (e,f) and $R_2$ (g,h). The orange lines are the 
analytically predicted phase boundaries, with solid and dashed lines 
denoting continuous and discontinuous transitions, respectively.}
\label{fig:validation}
\end{figure*}

\begin{figure*} [ht!]
\centering
\subfigure{\includegraphics[width=0.49\linewidth]{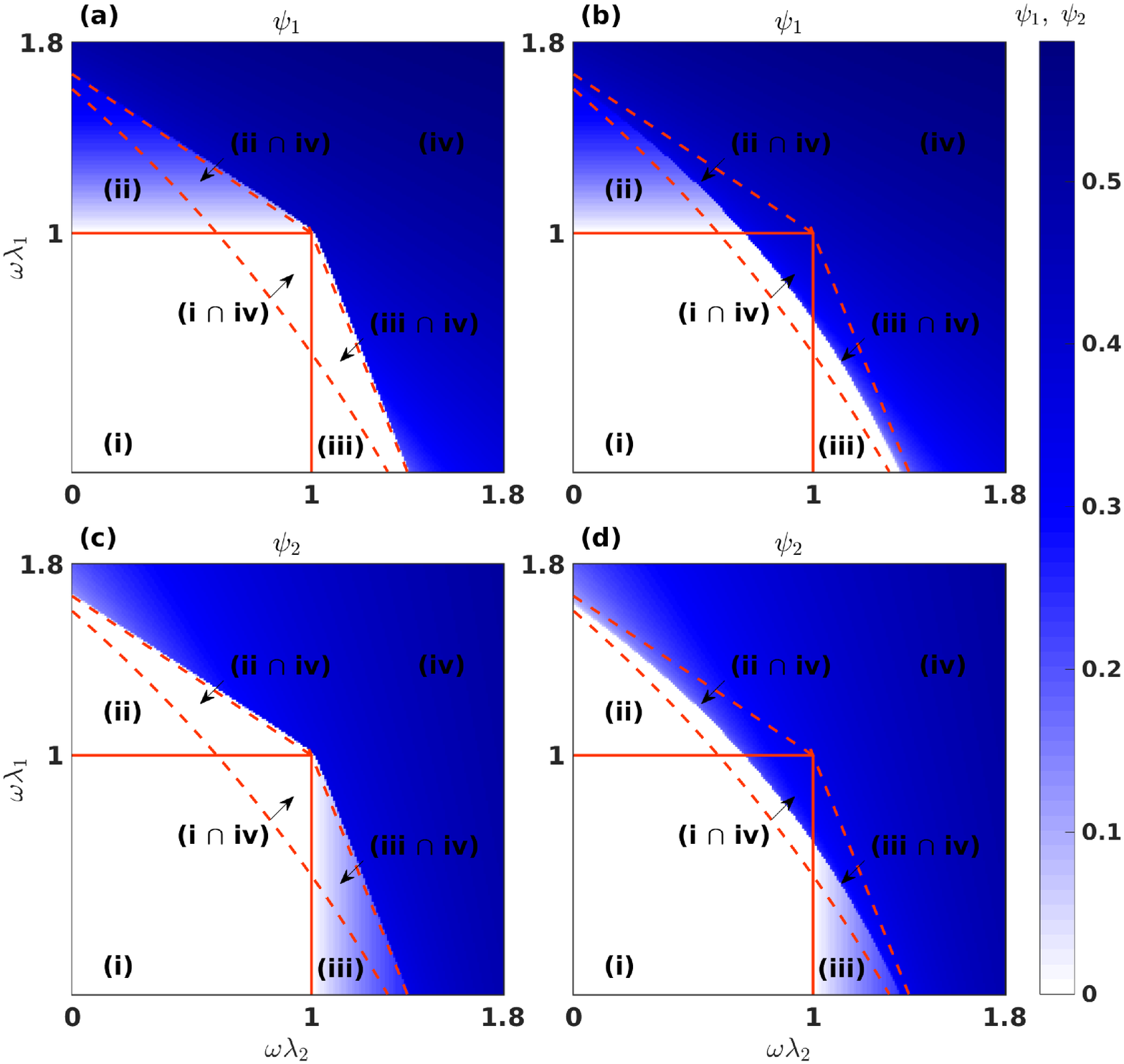}}
\subfigure{\includegraphics[width=0.49\linewidth]{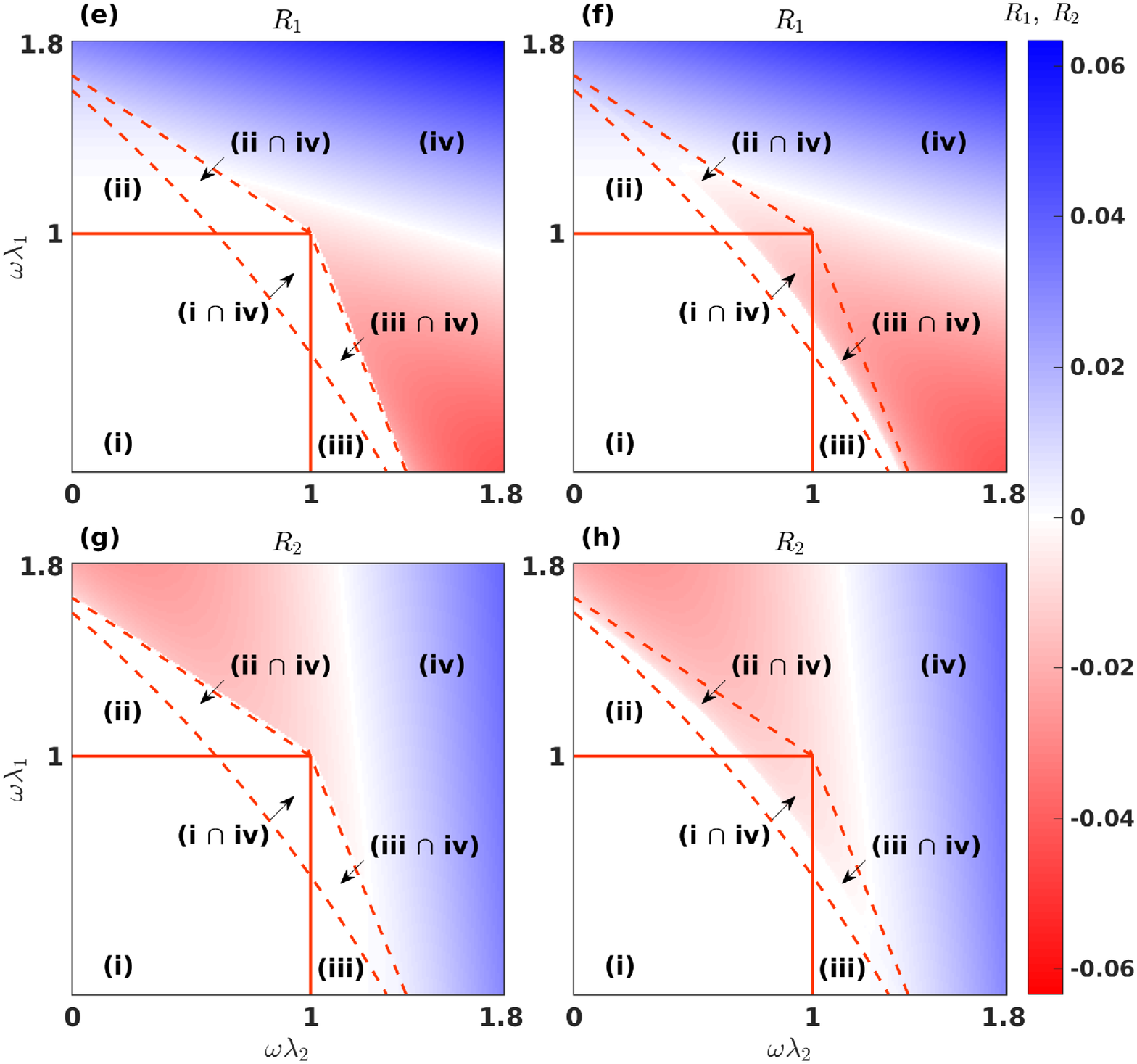}}
\caption{ \emph{Validation of analytic predictions for the power-law 
network with degree exponent $\beta=2.3$} The parameters of the spreading 
model are set as $\lambda_1^{\dagger}=3.5\omega^{-1}$ and 
$\lambda_2^{\dagger}=2.5\omega^{-1}$.}
\label{fig3}
\end{figure*}

\begin{figure*} [ht!]
\centering
\subfigure{\includegraphics[width=0.49\linewidth]{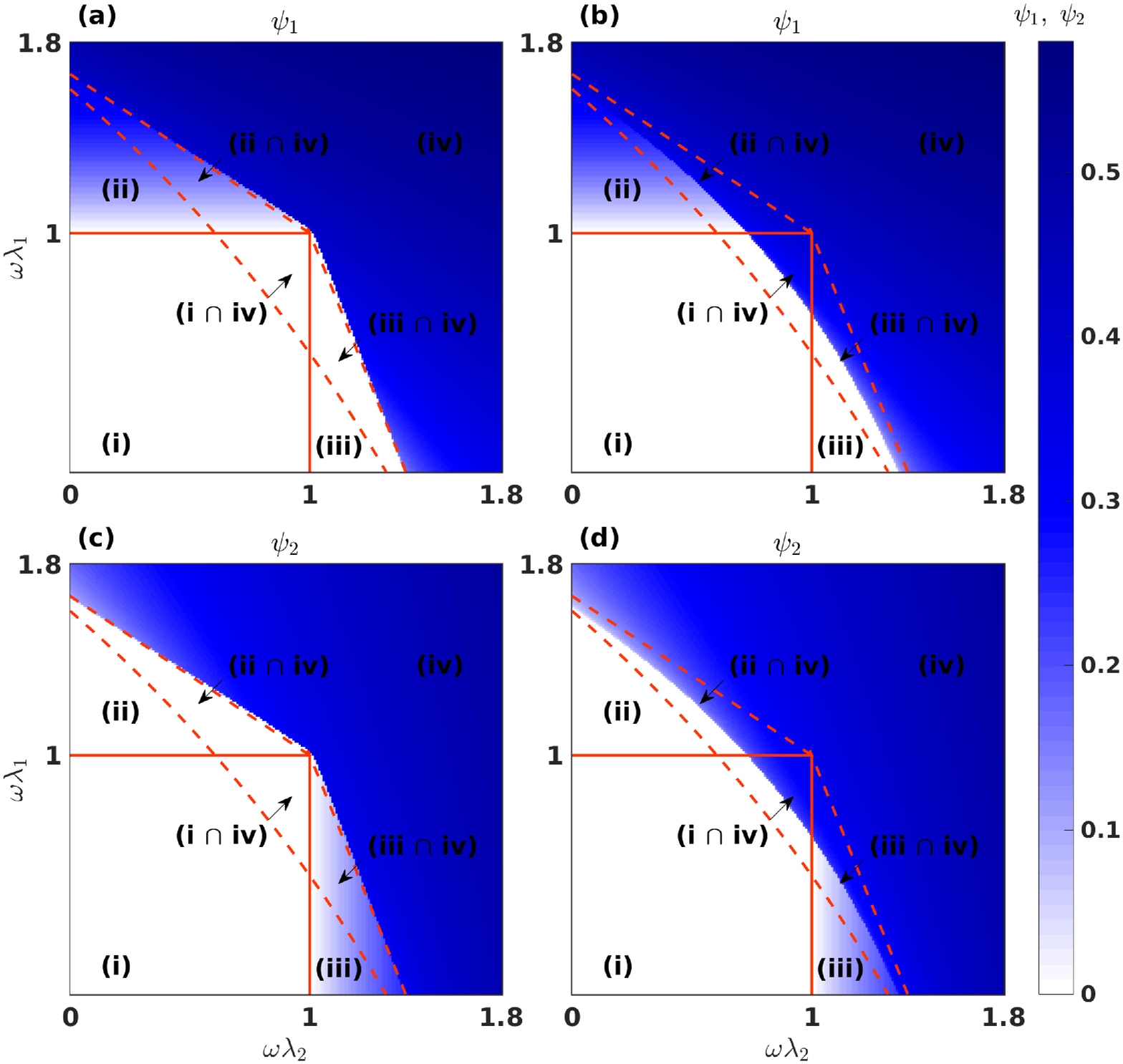}}
\subfigure{\includegraphics[width=0.49\linewidth]{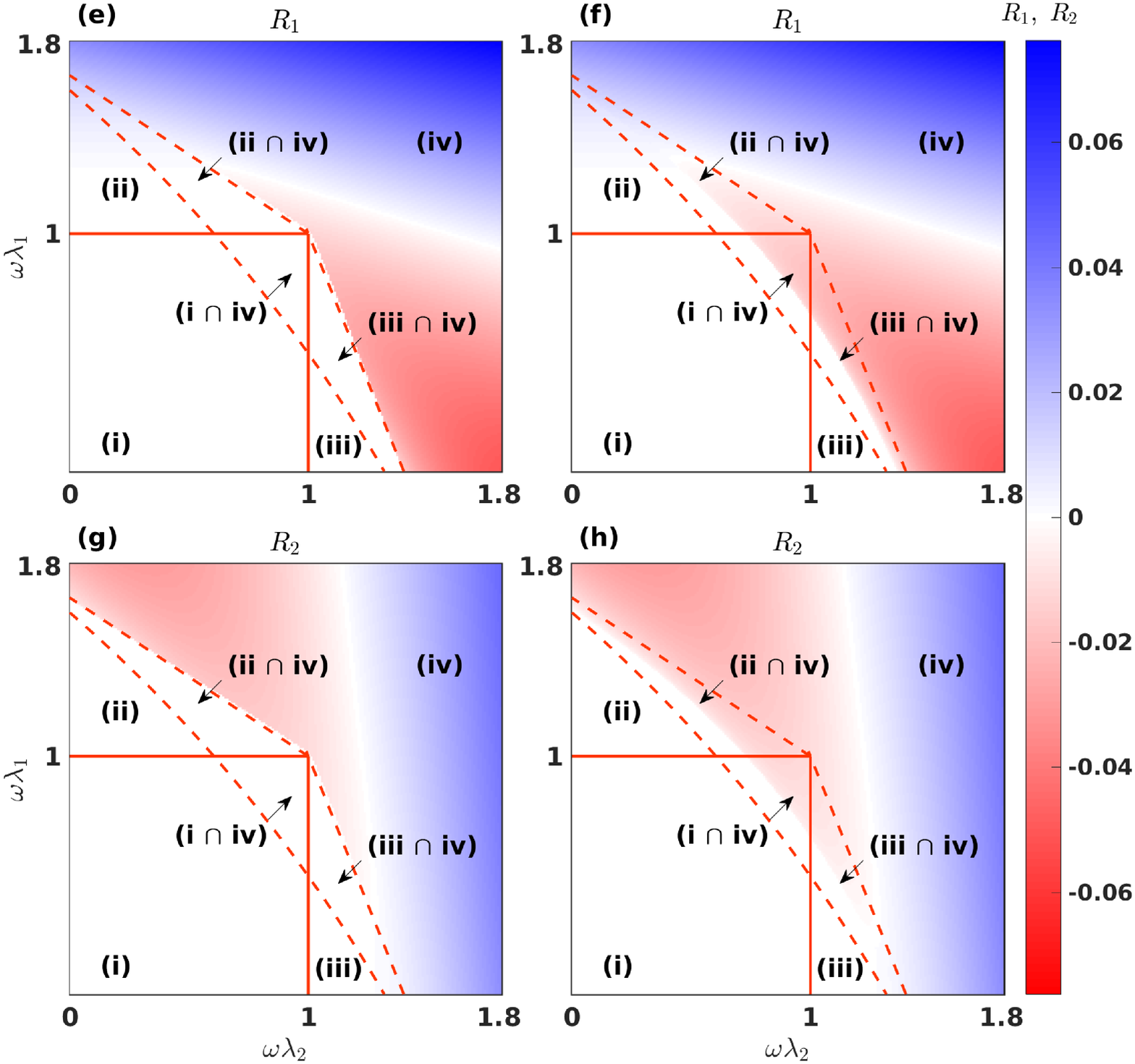}}
\caption{ \emph{Validation of analytic predictions for the power-law network 
with degree exponent $\beta=3$.} The parameters of the spreading model are 
set as $\lambda_1^{\dagger}=3.5\omega^{-1}$ and 
$\lambda_2^{\dagger}=2.5\omega^{-1}$.}
\label{fig4}
\end{figure*}

\begin{figure*} [ht!]
\centering
\subfigure{\includegraphics[width=0.49\linewidth]{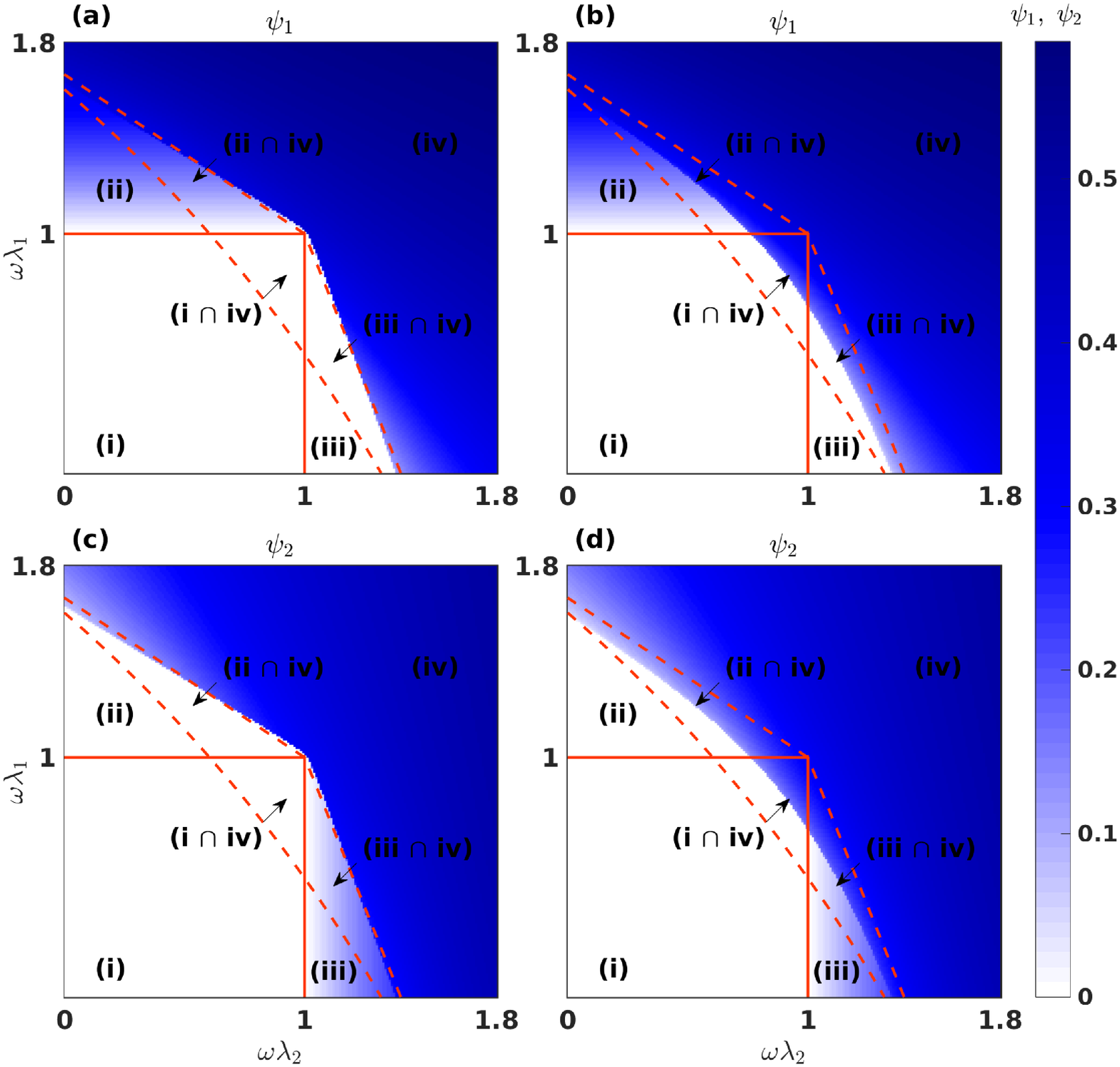}}
\subfigure{\includegraphics[width=0.49\linewidth]{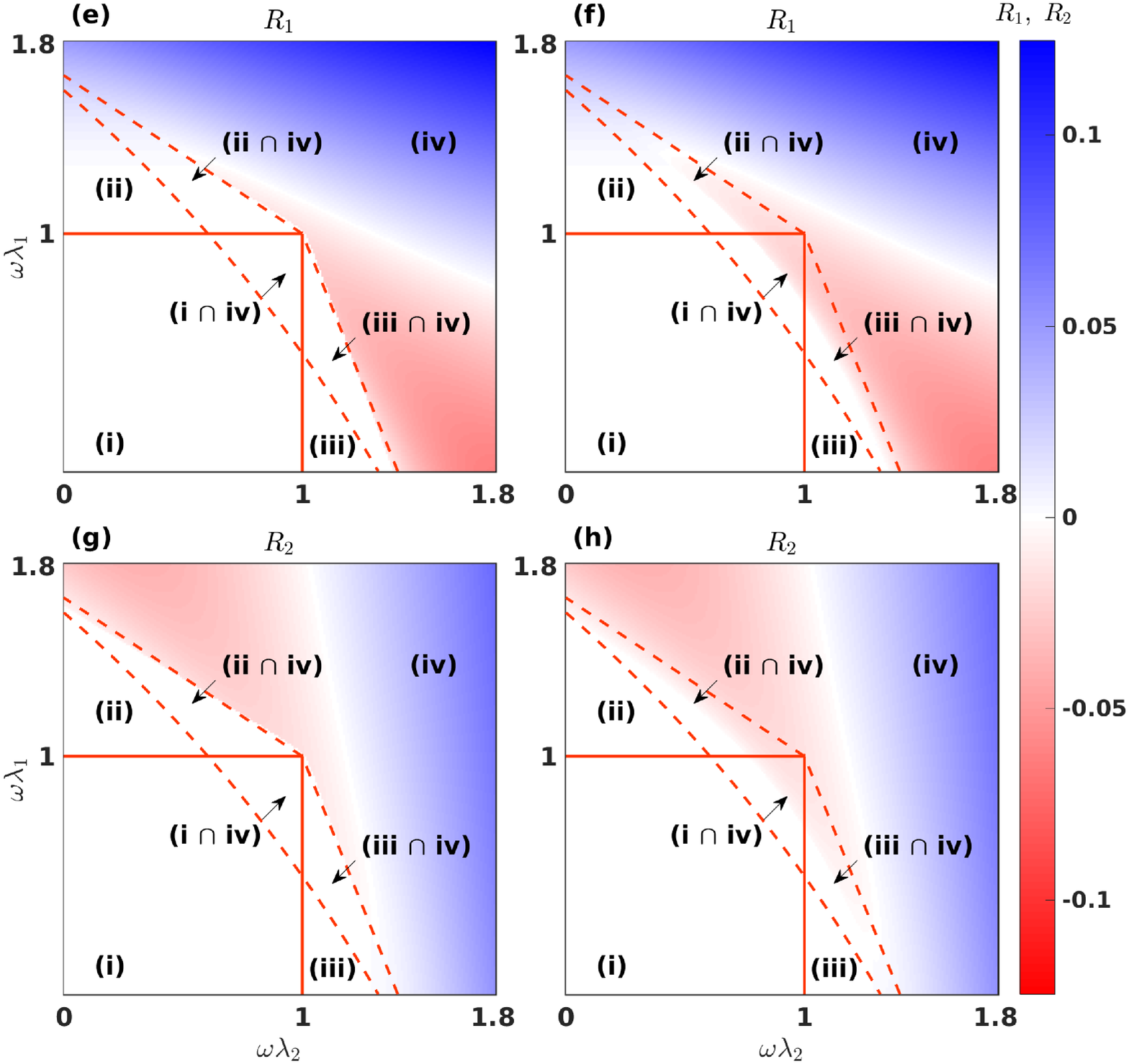}}
\caption{ \emph{Validation of analytic predictions for the power-law network 
with degree exponent $\beta=4$.} The parameters of the spreading model are 
set as $\lambda_1^{\dagger}=3.5\omega^{-1}$ and 
$\lambda_2^{\dagger}=2.5\omega^{-1}$.}
\label{fig5}
\end{figure*}

\begin{figure*} [ht!]
\centering
\subfigure{\includegraphics[width=0.49\linewidth]{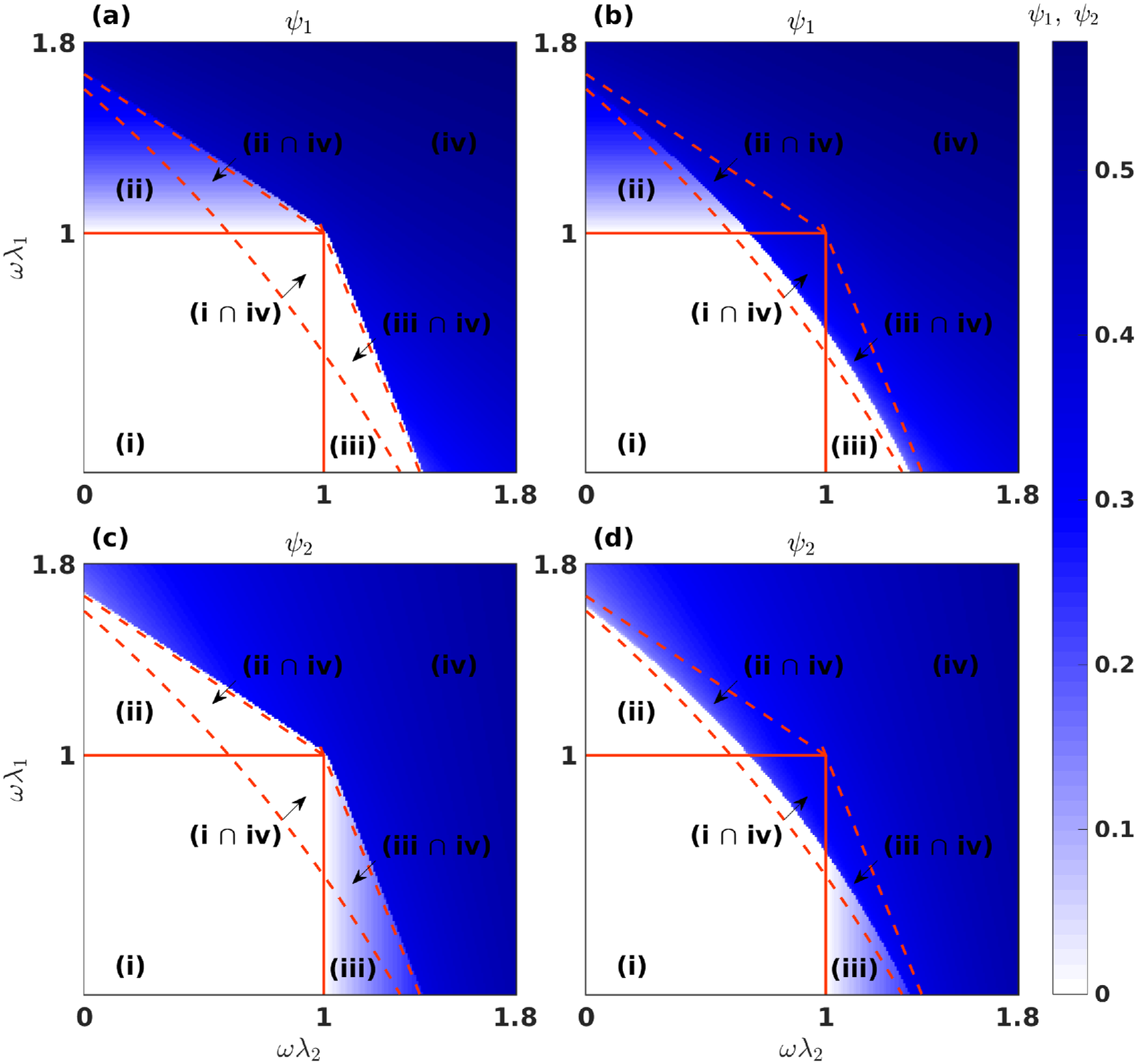}}
\subfigure{\includegraphics[width=0.49\linewidth]{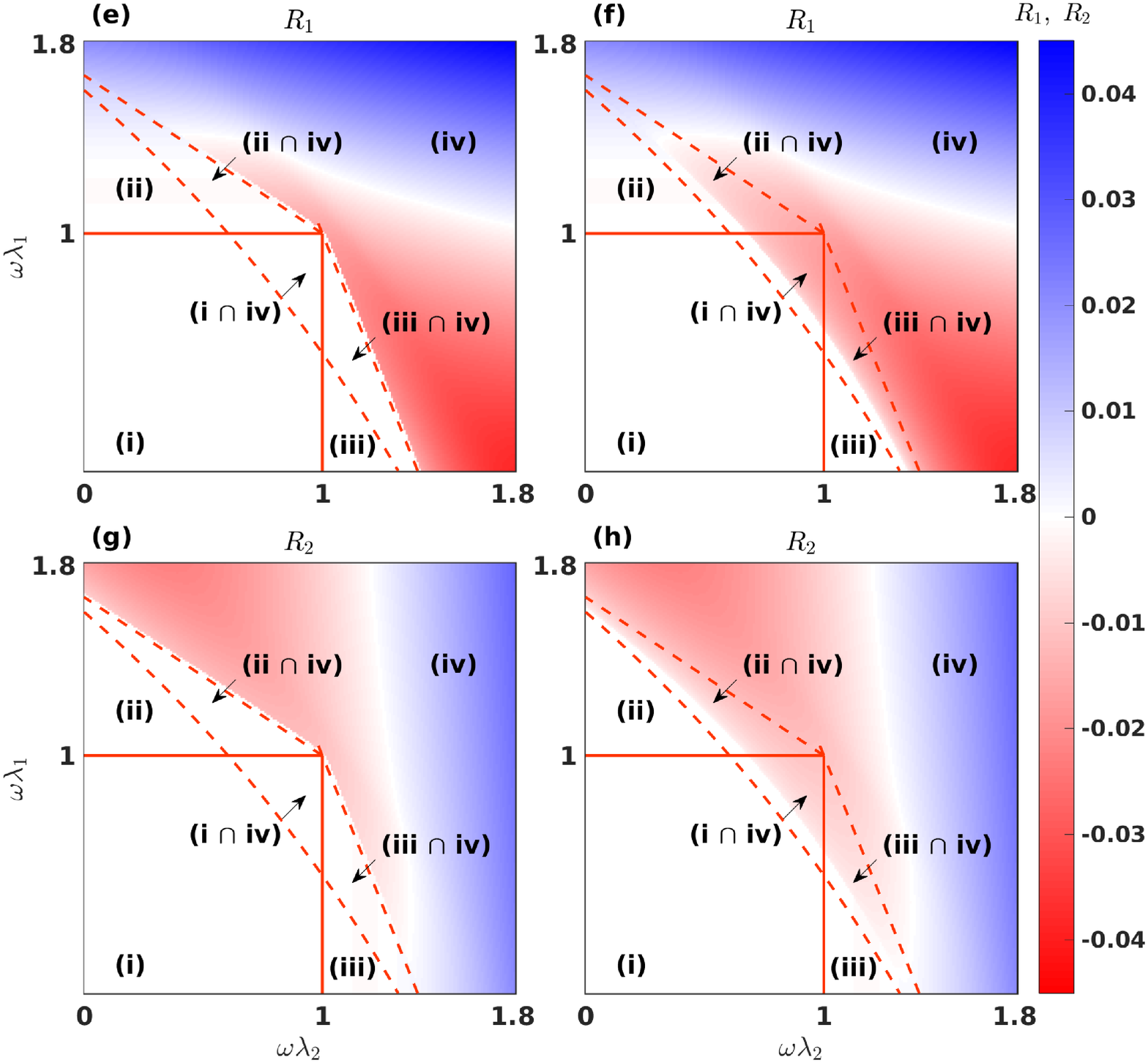}}
\caption{ \emph{Validation of analytic predictions for the Dolphins network.} 
The parameters of the spreading model are set as 
$\lambda_1^{\dagger}=3.5\omega^{-1}$ and $\lambda_2^{\dagger}=2.5\omega^{-1}$.}
\label{fig6}
\end{figure*}

\begin{figure*} [ht!]
\centering
\subfigure{\includegraphics[width=0.49\linewidth]{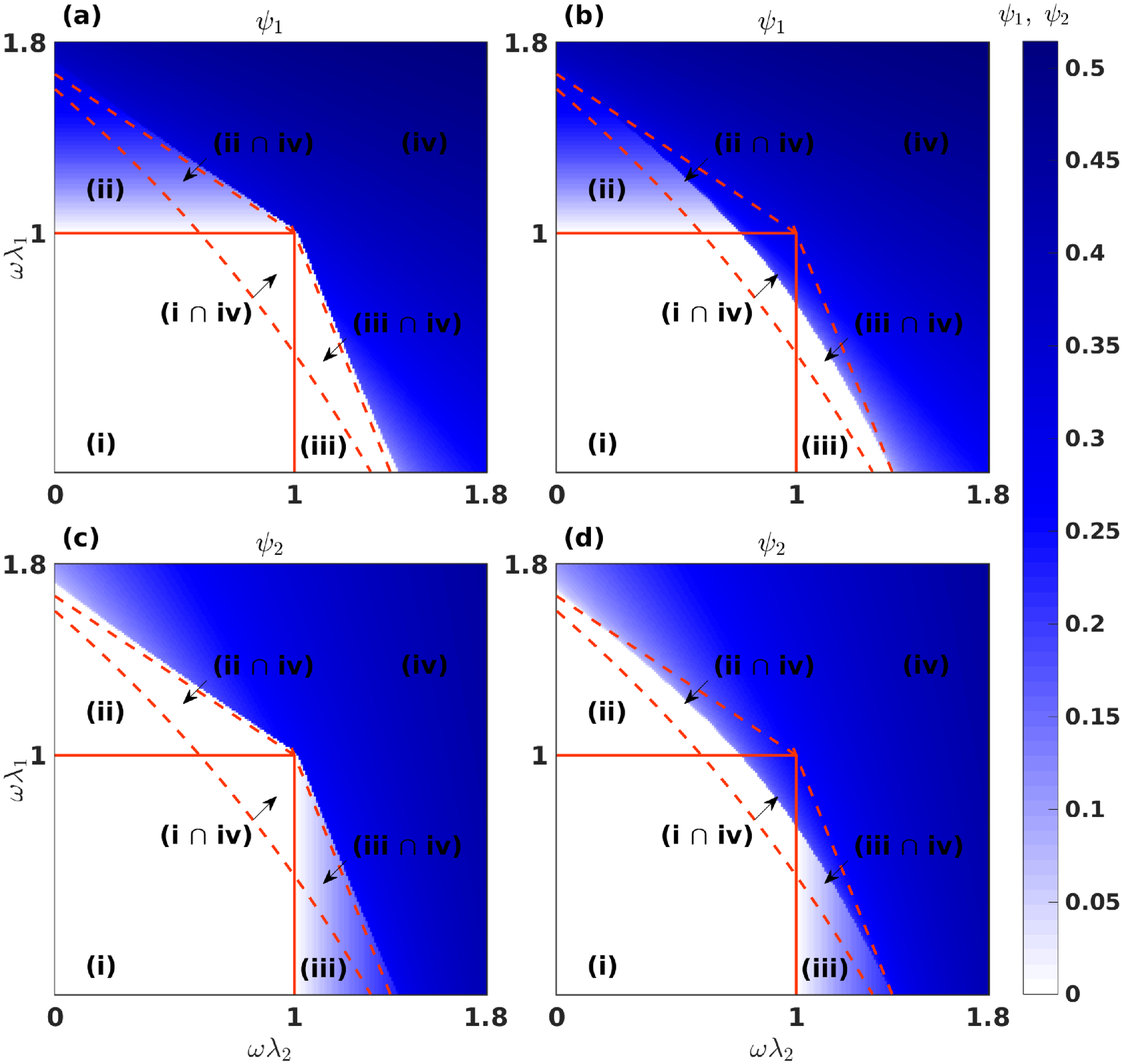}}
\subfigure{\includegraphics[width=0.49\linewidth]{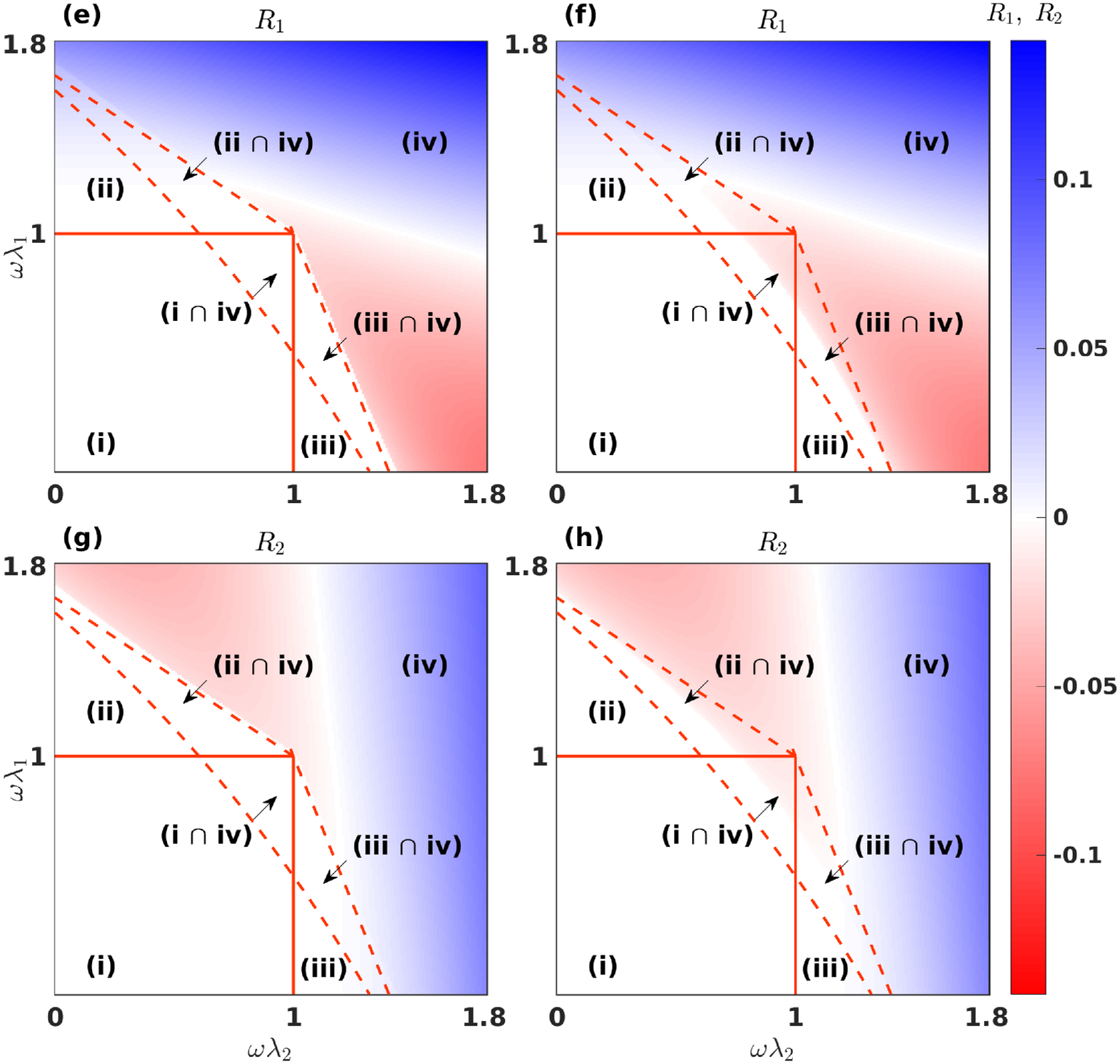}}
\caption{ \emph{Validation of analytic predictions for the HIV network.} 
The parameters of the spreading model are set as 
$\lambda_1^{\dagger}=3.5\omega^{-1}$ and $\lambda_2^{\dagger}=2.5\omega^{-1}$.}
\label{fig7}
\end{figure*}

\begin{figure*} [ht!]
\centering
\subfigure{\includegraphics[width=0.49\linewidth]{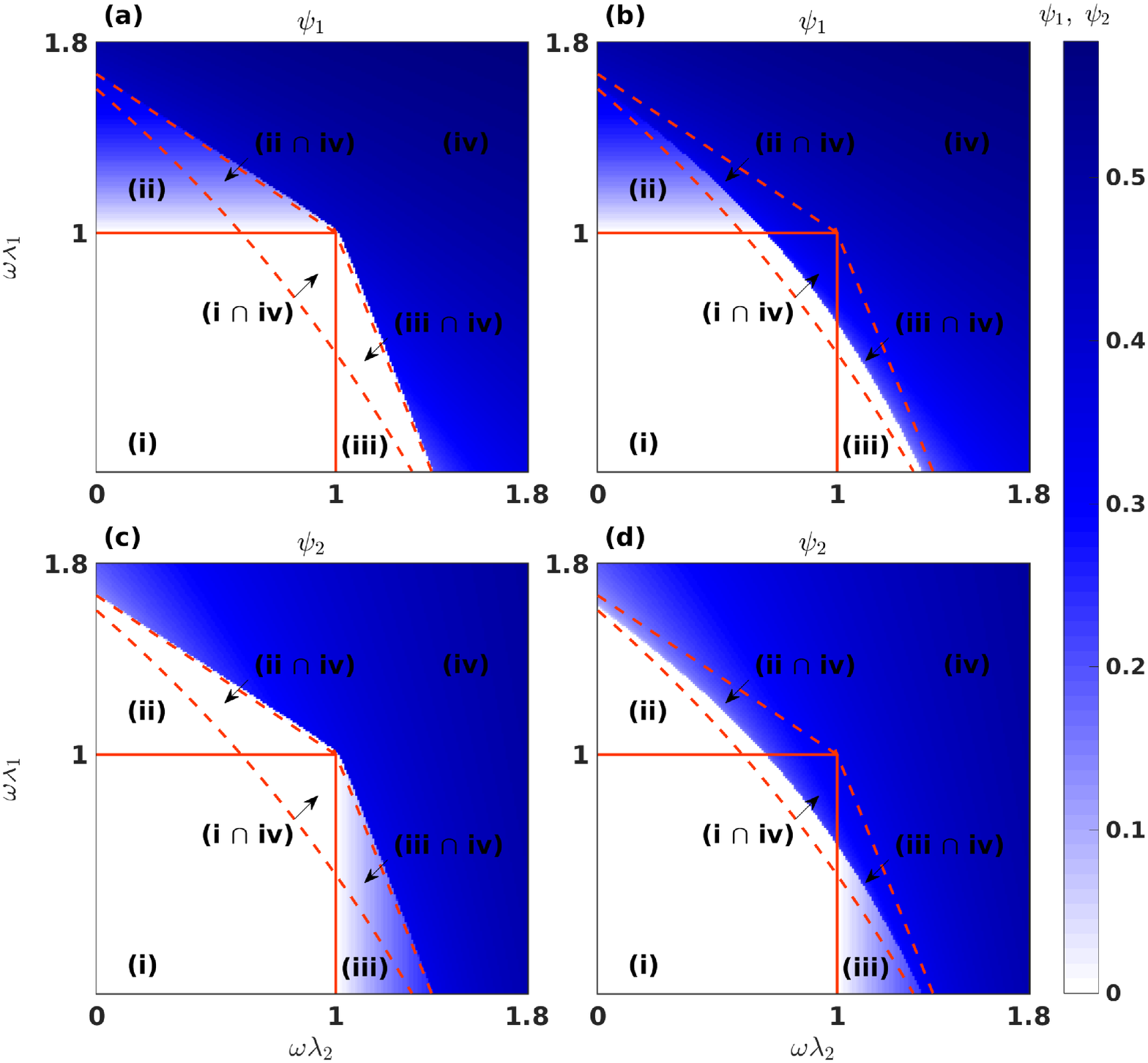}}
\subfigure{\includegraphics[width=0.49\linewidth]{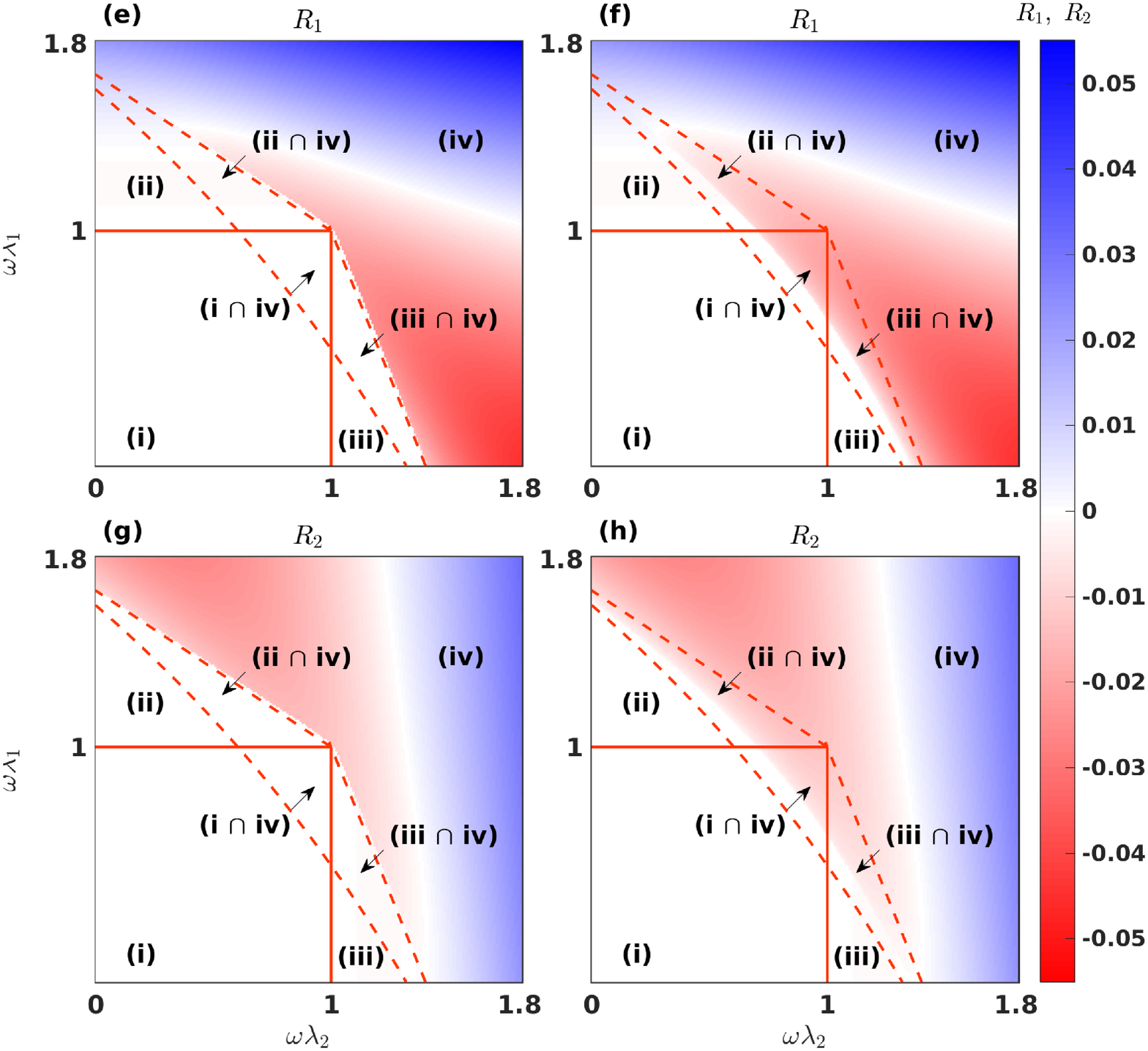}}
\caption{ \emph{Validation of analytic predictions for the Highschool network.}
The parameters of the spreading model are set as 
$\lambda_1^{\dagger}=3.5\omega^{-1}$ and $\lambda_2^{\dagger}=2.5\omega^{-1}$.}
\label{fig8}
\end{figure*}

\begin{figure*} [ht!]
\centering
\subfigure{\includegraphics[width=0.49\linewidth]{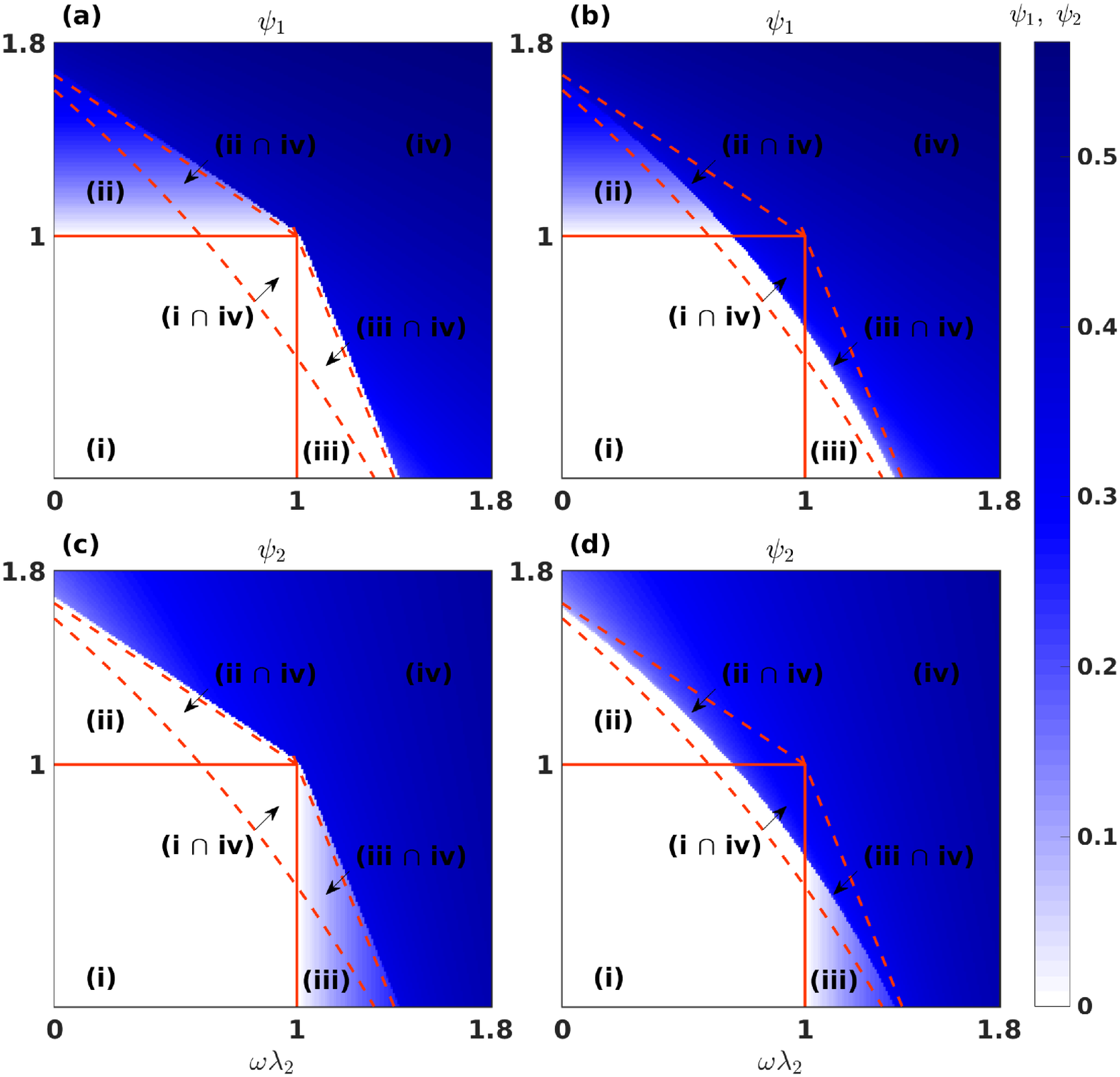}}
\subfigure{\includegraphics[width=0.49\linewidth]{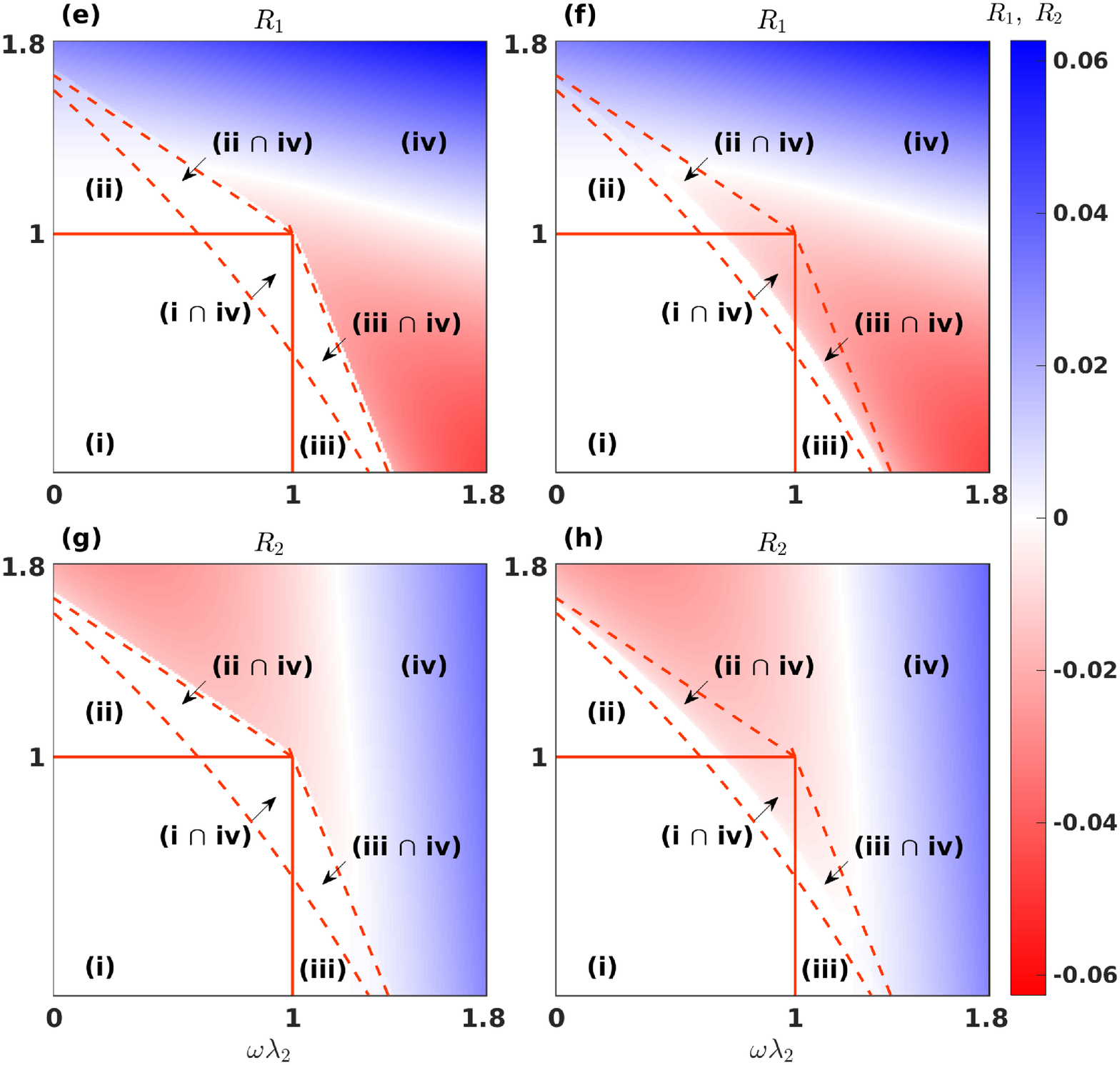}}
\caption{ \emph{Validation of analytic predictions for the Jazz network.} 
The parameters of the spreading model are set as 
$\lambda_1^{\dagger}=3.5\omega^{-1}$ and $\lambda_2^{\dagger}=2.5\omega^{-1}$.}
\label{fig9}
\end{figure*}

We provide a numerical illustration of the analytic prediction on the
interplay between discontinuous transitions and hystereses
with an Erd\H{o}s-R\'{e}nyi graph of size $N=100$ and average degree
$\langle k \rangle=4$. The inequality (\ref{eq:hysteresisCond}) divides
the $\lambda_1^{\dagger}$-$\lambda_2^{\dagger}$ plane into two regions,
as shown in Fig.~\ref{fig:PD}(a). Above the curve defined by
\begin{displaymath}
\lambda^{\dagger}_1+\lambda^{\dagger}_2-\omega \lambda^{\dagger}_1\lambda^{\dagger}_2=0,
\end{displaymath}	
a hysteresis region appears while it is absent below. In the limit
$\lambda_a^{\dagger}\to \infty$, the curve approaches
$\lambda_b^{\dagger}=\omega^{-1}$, as shown by the orange dashed lines in
Fig.~\ref{fig:PD}(a). Note that the curve avoids the dashed lines for
finite $\lambda_a^{\dagger}$. Since $\omega^{-1}$ is also the outbreak
threshold of the classic SIS model for a single epidemic, a necessary
condition for a hysteresis is that $\lambda_a^{\dagger}$ must be larger
than the classic threshold. A special case is
$\lambda_1^{\dagger}=\lambda_2^{\dagger}$, where \eqref{eq:hysteresisCond}
implies that, for a hysteresis to arise, the inequality
$\lambda_1^{\dagger}=\lambda_2^{\dagger}>2\omega^{-1}$ must hold.
That is, the interactive transmission rate must at least twice the classic
SIS threshold for a hysteresis to arise, suggesting that networks with a
larger leading eigenvalue are more prone to hystereses.
Two representative phase diagrams in the $\lambda_1$-$\lambda_2$ plane with
fixed values of $\lambda_1^{\dagger}$ and $\lambda_2^{\dagger}$ are shown in
Figs.~\ref{fig:PD}(b) and \ref{fig:PD}(c), corresponding to the points $b$
and $c$ in Fig.~\ref{fig:PD}(a), respectively. For point $b$, no hysteresis
can arise for any values of ($\lambda_1$, $\lambda_2$) and the phase
transitions between different neighboring phase regions are continuous, as
indicated by the solid lines in Fig.~\ref{fig:PD}(b). For point $c$
that is slightly above the hysteresis boundary, region {\em (\romannumeral4)}
overlaps with regions {\em (\romannumeral1)}, {\em (\romannumeral2)} and
{\em (\romannumeral3)}, where a hysteresis can arise. Crossing into
region {\em (\romannumeral4)} from any one of the phase regions
({\em \romannumeral1} $\ \cap$ {\em \romannumeral4}),
({\em \romannumeral2} $\ \cap$ {\em \romannumeral4}) and
({\em \romannumeral3} $\ \cap$ {\em \romannumeral4}),
a discontinuous outbreak transition occurs with some $\psi_a$ changing
abruptly from zero to a nonzero value. Along the path
({\em \romannumeral1} $\ \cap$ {\em \romannumeral4})$\ \to$
({\em \romannumeral1}), ({\em \romannumeral2} $\ \cap$ {\em \romannumeral4})$
\ \to$ ({\em \romannumeral2}) and ({\em \romannumeral3} $\ \cap$
{\em \romannumeral4})$\ \to$ {\em (\romannumeral3}), the system displays a
discontinuous transition to extinction at which at least one epidemic changes
abruptly from a nonzero value to zero. All the phase boundaries with
discontinuous transitions are indicated by the dashed lines in
Fig.~\ref{fig:PD}(c), where the two tricritical points separating continuous
from discontinuous transitions are marked (white circles).

Are the phase diagrams obtained from the reduced mean field equations
accurate in comparison with those from the original mean field equations?
In the presence of the fluctuation terms $R_{a}$,
Eqs.~\eqref{eq:reducedQMF} are exactly equivalent to Eqs.~\eqref{eq:QMF}.
Consider a system of dimension $2N+2$, which consists of Eqs.~\eqref{eq:QMF}
and Eqs.~\eqref{eq:reducedQMF}. A stable equilibrium point
$\left(p_{1,i},p_{2,i}\right)_{1\leq i\leq N}$ of the subsystem
Eqs.~\eqref{eq:QMF} is also one for the $2N+2$ system with
$\psi_a=\alpha^T p_{a}$. Consider a stable equilibrium point with which
neither epidemic has an outbreak. Substituting
$p_{a,1}=\cdots=p_{a,N}=0$ and $\psi_a=0$ into the remainder term
(with full expression in Sec.~\ref{sec:SDR}), we have
$R_{a}=0$ for $a\in\{1,2\}$. In this case the remainder terms can be
ignored. Since a zero stable equilibrium point of Eqs.~\eqref{eq:QMF}
implies the existence of exactly such a point of Eqs.~\eqref{eq:reducedQMF}
(with no remainder terms) and vice versa, any outbreak transition threshold
from phase (\romannumeral1) is expected to be exact for Eqs.~\eqref{eq:QMF}.

There are two cases where the remainder terms do not vanish and can lead
to inaccuracies of the analytic prediction. The first case is when
Eqs.~\eqref{eq:QMF} exhibit a stable equilibrium point at which there
is an outbreak for epidemic $1$ but extinction for epidemic $2$:
$R_{1}=0$ and $R_{2}\neq 0$. The second case is when Eqs.~\eqref{eq:QMF}
have a stable equilibrium point with an outbreak for both epidemics:
$R_{a}\neq 0$ for $a\in\{1,2\}$. Since the remainder terms are small by
construction, they lead to corrections that can be neglected, which have
been verified numerically. Especially, for the Erd\H{o}s-R\'{e}nyi network
in Fig.~\ref{fig:PD}, we solve Eqs.~\eqref{eq:QMF} numerically and compare
the solutions with the analytic phase diagram obtained from
Eqs.~\eqref{eq:reducedQMF}. The values of $\psi_a$ obtained from
Eqs.~\eqref{eq:QMF} in the $\lambda_1$-$\lambda_2$ plane are shown in
Figs.~\ref{fig:validation}(a-d), for $\lambda_1^{\dagger}=3.5\omega^{-1}$
and $\lambda_2^{\dagger}=2.5\omega^{-1}$ [so that \eqref{eq:hysteresisCond}
is satisfied, guaranteeing a hysteresis]. Since in the hysteresis region there
are two stable equilibrium points for each $\psi_a$, we plot separately the
two solutions for $\psi_{1}$ in Figs.~\ref{fig:validation}(a) and
\ref{fig:validation}(b), and those for $\psi_{2}$ in
Figs.~\ref{fig:validation}(c) and \ref{fig:validation}(d), respectively. The
phase diagram from original Eqs.~\eqref{eq:reducedQMF} is also shown in
Fig.~\ref{fig:validation} by the orange solid and dashed lines for
continuous and discontinuous transitions, respectively. Our analytical
phase diagram predicts accurately all outbreak transitions:
({\em \romannumeral1})$\ \to$ ({\em \romannumeral2}), ({\em \romannumeral1})$
\ \to$
({\em \romannumeral3}), ({\em \romannumeral1} $\ \cap$ {\em \romannumeral4})$
\ \to$
({\em \romannumeral4}), ({\em \romannumeral2} $\ \cap$ {\em \romannumeral4})$
\ \to$
({\em \romannumeral4}) and ({\em \romannumeral3} $\ \cap$ {\em \romannumeral4})$
\ \to$
({\em \romannumeral4}).
However, quantitatively, the predicted extinction transitions
({\em \romannumeral1} $\ \cap$ {\em \romannumeral4})$\ \to$
({\em \romannumeral1}), ({\em \romannumeral2} $\ \cap$ {\em \romannumeral4})$
\ \to$ ({\em \romannumeral2}) and ({\em \romannumeral3} $\ \cap$
{\em \romannumeral4})$\ \to$ ({\em \romannumeral3}) are
less accurate, due to the nonzero remainders $R_1$ and $R_2$ as a result of
the loss of stability of an equilibrium point with an outbreak for both
epidemics. The values of the remainders $R_1$ and $R_2$ at equilibrium are
shown in Figs.~\ref{fig:validation}(e-h). The value of $R_1$ for the two
solutions of $\psi_1$ are shown in Figs.~\ref{fig:validation}(e) and
\ref{fig:validation}(f), respectively. Similarly, $R_2$ for the two solutions
of $\psi_2$ are shown in Figs.~\ref{fig:validation}(g) and
\ref{fig:validation}(h), respectively. The predictions are qualitatively
correct.

Next we consider tests and validation of our analytic prediction from 
Eqs.~\eqref{eq:reducedQMF} for a variety of networks, including synthetic 
networks with strong and weak degree heterogeneity, and real-world networks. 
For synthetic networks, we have already shown the results for an ER network.
Here we also consider networks generated from the uncorrelated configuration 
model (UCM) with a power-law degree distribution $p(k)\sim k^{-\beta}$.
Specifically, we consider three networks with different degree exponents: 
(1) PL-2.3 with $\beta=2.3$, (2) PL-3 with $\beta=3$ and (3) PL-4 with 
$\beta=4$. For empirical networks, we have 
(4) \emph{Dolphins}~\cite{dolphins}, a social network of bottle-nose dolphins;
(5) \emph{HIV}~\cite{hiv}, a network of sexual contacts between people
involved in the early spread of the human immunodeficiency virus (HIV);
(6) \emph{Highschool}~\cite{highschool}, a friendship network between
boys in a small high school, and (7) \emph{Jazz}~\cite{jazz}, a collaboration 
network between Jazz musicians. The networks are downloaded from 
Ref.~\cite{konect}. 

Basic features and parameters of the networks considered are listed in 
Table~\ref{prop}. Note that \emph{Highschool} is a directed and weighted 
network. Here we simply take it as undirected by assuming that there is an 
undirected edge between node $i$ and $j$ if there is at least a directed 
edge between the two nodes in either direction, with the edge weights ignored. 

\begin{table}[htbp]
\caption{Basic topological features of seven real networks: $N$ and $M$ 
are the number of nodes and edges, respectively, $C$ is the clustering 
coefficient \cite{watts1998}, $r$ is the assortative 
coefficient~\cite{newman2002}, $\langle k\rangle$ is the average degree, 
$H$ is the degree heterogeneity which defined as 
$H={\langle k^2\rangle}/{{\langle k\rangle}^2}$, and 
$\langle d\rangle$ is the average shortest distance.}

\begin{tabular}{ccccccccc}
\toprule
\hline
\hline
  & $N$ & $M$ & $C$ & $r$ & $k_{\mathrm{max}}$ & $\langle k\rangle$ & $H$ & $\langle d\rangle$\\
\midrule
\hline
\emph{ER} & 100 & 200 & 0.025 & 0.027 & 9 & 4 & 1.228 & 3.436\\
\emph{PL-2.3} & 100 & 234 & 0.040 & -0.077 & 10 & 4.680 & 1.162 & 3.095\\
\emph{PL-3} & 100 & 216 & 0.042 & -0.005 & 10 & 4.320 & 1.164 & 3.239\\
\emph{PL-4} & 100 & 185 & 0.033 & 0.022 & 10 & 3.700 & 1.108 & 3.709\\
\emph{Dolphins} & 62 & 159 & 0.259 & -0.043 & 12 & 5.130 & 1.327 & 3.357\\
\emph{HIV} & 40 & 41 & 0.042 & -0.279 & 8 & 2.050 & 1.512 & 4.474\\
\emph{Highschool} & 70 & 274 & 0.465 & 0.083 & 19 & 7.829 & 1.190 & 2.640\\
\emph{Jazz} & 198 & 2742 & 0.618 & 0.021 & 100 & 27.697 & 1.396 & 2.236 \\
\bottomrule
\hline
\hline
\end{tabular}
\label{prop}
\end{table}

For all the networks, we set $\lambda_1^{\dagger}=3.5\omega^{-1}$ and
$\lambda_2^{\dagger}=2.5\omega^{-1}$ to guarantee the emergence of a
hysteresis region. To have an idea of the size of the correction terms 
$R_a$, we show the values of $R_a$ at equilibrium. The results of (1) 
\emph{PL-2.3} (2) \emph{PL-3}, (3) \emph{PL-4}, (4) \emph{Dolphins},
(5) \emph{HIV}, (6) \emph{Highschool} and (7) \emph{Jazz} are shown in 
Figs.~(\ref{fig3}),~(\ref{fig4}),~(\ref{fig5}), (\ref{fig6}),~(\ref{fig7}),
~(\ref{fig8}) and~(\ref{fig9}), respectively. In each figure, subfigures (a) 
and (b) correspond to the values of $\psi_1$, while (e) and (f) are the 
corresponding values of $R_1$. Similarly, (c) and (d) correspond to the 
values of $\psi_2$, while (g) and (h) are the corresponding values of $R_2$.
We see that, for all the networks tested, the analytic phase diagram 
predicts quantitatively and accurately the outbreak transitions, while the 
predicted extinction transitions are qualitatively correct. The values of 
correction terms $R_a$ are near zero for the outbreak transitions, while 
have relatively larger magnitudes near extinction transitions.

\section{Discussion} \label{sec:discussion}

We have analytically predicted the phase diagram of interacting SIS spreading 
dynamics using the technique of spectral dimension reduction and provided 
numerical validation. 
The analytic phase diagram elucidates the interplay between discontinuous 
transitions and hystereses as well as the emergence of tricritical points.
This method can also be applied to study other interacting epidemic models. 
For general epidemic models, a one-dimensional description
of each epidemics is not sufficient~\cite{laurence2019spectral}. Determining
the number of macroscopic observables required for general epidemic models
needs further exploration.

Previous theoretical methods for interacting spreading dynamics such as
QMF theory~\cite{wang2003epidemic} employ $2N$ equations, where $N$ is the
network size. For large networks, it is computationally demanding to solve
the equations to determine the stability of the fixed points, as this requires
manipulating the $2N\times 2N$ Jacobian matrix. It is thus infeasible to use
the QMF to map out the phase diagram for interacting spreading dynamics on
complex networks, preventing us from gaining a full understanding of the
interplay between network topology and the spreading dynamical process as a
full phase diagram would reveal. The same difficulty arises with a naive
application of the SDR method~\cite{laurence2019spectral} in order to obtain
the phase diagram for interacting spreading dynamics. In contrast,
our approach gives a full picture of the phase diagram on large complex
networks with an arbitrary topology through an effective two-dimensional
system, revealing rich phenomena that have not been systemically investigated.
While many previous studies employed the mean-field theory to study different
types of spreading dynamics on complex
networks~\cite{PSV:2001,BPSV:2003,CPS:2006}, our work is not a simple
application of the mean-field theory. In fact, we go way beyond by
obtaining, for the first time to our knowledge, a global phase diagram laying
out a clear picture of all possible dynamical states and the transitions among
them through a comprehensive stability analysis - both at an unprecedented
level of details.

Taken together, our work gives a full picture of the dependence of phase
transition on network topology and spreading parameters for SIS dynamics,
and thus lays a foundation for intervening or harnessing this type of 
interacting spreading processes. For instance, our phase diagram gives possible
routes for controlling the type of phase transition through perturbations to
the network structure or for controlling one spreading process through
manipulating another interacting process. It should be cautioned that,
while the SIS model provides phenomenological insights into relatively
simply spreading processes and is thus a conceptually useful paradigmatic
model, it may be too simplistic to describe spreading processes in the real
world which can be significantly more complicated. To apply our analytic
approach to irreversible epidemic processes beyond the SIS dynamics is
possible but remains to be studied.

\section*{Acknowledgments}

This work was partially supported by NNSF of China under Grants Nos.~61903266,
61433014 and 61673086), China Postdoctoral Science Foundation under Grant
No.~2018M631073, China Postdoctoral Science Special Foundation under Grant
No.~2019T120829), the Science Strength Promotion Program of the University
of Electronic Science and Technology of China under Grant No.~Y030190261010020,
and Fundamental Research Funds for the Central Universities. YCL is supported
by ONR under Grant No. N00014-16-1-2828.

\subsection*{Appendix A: Proof for $\mu \geq 1$}
\setcounter{equation}{0}
\renewcommand\theequation{A.\arabic{equation}}

To prove $\mu\geq 1$, we rewrite Eq.~\eqref{eq:muFormula} as
\begin{equation}
\mu^{-1}=\frac{\alpha^T G \alpha}{\alpha^T K \alpha}.
\end{equation}
Since $K$ is positive definite, it can be decomposed as $K=K^{1/2}K^{1/2}$,
where $K^{1/2}$ is a diagonal matrix whose entries are the square root of
the degrees. Let $y=K^{1/2}\alpha$. We have $\alpha=K^{-1/2}y$. Substituting
this back to $\mu^{-1}$ gives
\begin{equation}
\mu^{-1}=\frac{y K^{-1/2} G K^{-1/2} y}{y^T y},
\end{equation}
which is the Rayleigh quotient of matrix $K^{-1/2} G K^{-1/2}$ and, hence,
we have $\mu^{-1}\leq \delta_1 $, where $\delta_1$ is the largest eigenvalue
of the matrix $K^{-1/2} G K^{-1/2}$. Recall that the symmetric normalized
Laplacian matrix of $G$ is defined as
\begin{equation}
L^{\mathrm{sym}}=I-K^{-1/2} G K^{-1/2},
\end{equation}
which has a smallest eigenvalue $\zeta_n=0$. As a result, we have
$\delta_1=1-\zeta_n=1$, which gives $\mu\geq 1$.


%
\end{document}